\documentclass[12pt, a4paper]{article}
\pdfoutput=1

\usepackage{amsmath}
\usepackage{amsfonts}
\usepackage{amssymb}
\usepackage{latexsym}
\usepackage{graphicx}
\usepackage{color}
\usepackage{mathrsfs}
\usepackage{slashed}
\usepackage{hyperref}
\usepackage{datetime}
\usepackage{enumerate}

\numberwithin{equation}{section}

\hypersetup{colorlinks, citecolor=bluscuro, linkcolor=bluscuro, urlcolor=bluscuro}
\definecolor{rossos}{rgb}{0.8,0.2,0.3}
\definecolor{bluscuro}{rgb}{0.15, 0.2, 0.9}
\definecolor{verdes}{rgb}{0.1, 0.5, 0.1}

\makeatother   % Cancel the effect of \makeatletter
\newcommand{\GeV}{{\rm \,GeV}}
\newcommand{\TeV}{{\rm \,TeV}}

\newcommand{\lp}{\left(}
\newcommand{\rp}{\right)}

\newcommand{\pslash}{\!\not\! p}

\newcommand{\eq}[1]{Eq.~(\ref{#1})}
 \def\be   {\begin{equation}}   \def\ee   {\end{equation}}
 \def\ba   {\begin{array}}      \def\ea   {\end{array}}
 \def\bea  {\begin{eqnarray}}   \def\eea  {\end{eqnarray}}
 \def\bean {\begin{eqnarray*}}  \def\eean {\end{eqnarray*}}

\begin{document}

\def\thefootnote{\arabic{footnote}}
%\setcounter{footnote}{0}
%\pagestyle{empty}
%
%\newpage
%\pagestyle{plain}
%\setcounter{page}{1}

\begin{flushright}
\small
LPTENS--12/12\\
L2C:12-032
\end{flushright}

\vspace{.3in}

\begin{center}
{\LARGE \bf Composite Scalar Dark Matter}\\
\vspace{0.3cm}
%{\LARGE \bf with a composite Higgs}

\vspace{2 cm}

{\large Michele Frigerio\,$^a$, Alex Pomarol\,$^b$, Francesco Riva\,$^c$ and  Alfredo Urbano\,$^d$}

\vspace{1cm}

\centerline{$^a$ {\it  CNRS, Laboratoire Charles Coulomb, UMR 5221, F-34095 Montpellier, FRANCE} \&}
\centerline{\it   Universit\'e Montpellier 2, Laboratoire Charles Coulomb, UMR 5221, F-34095 Montpellier, FRANCE}
\vspace{0.2cm}
\centerline{$^b${\it  Departament de Fisica, Universitat Aut\`onoma de Barcelona, 08193 Bellaterra, Barcelona, SPAIN}}
\vspace{0.2cm}
\centerline{$^c${\it  IFAE, Universitat Aut\`onoma de Barcelona, 08193 
Bellaterra, Barcelona, SPAIN}}
\vspace{0.2cm}
\centerline{$^d${\it  Laboratoire de Physique Th\'eorique de l'\'Ecole Normale Sup\'erieure,}}
\centerline{{\it  24 rue Lhomond, F-75231 Paris, FRANCE}}

\end{center}
\vspace{.8cm}

\begin{abstract}

We show that  the dark matter  (DM) could be a light composite scalar $\eta$,
emerging from a TeV-scale strongly-coupled sector as a pseudo Nambu-Goldstone boson (pNGB). 
Such state arises naturally  in scenarios where the Higgs is also a composite pNGB,
as in  $O(6)/O(5)$ models, which  are particularly predictive, since the low-energy  interactions of $\eta$ are determined by  symmetry considerations. We identify the region of parameters where $\eta$ has the required DM relic density,
satisfying at the same time the constraints from Higgs searches at the LHC, as well as DM direct searches. Compositeness, in addition to justify the lightness of the scalars, can enhance  the DM scattering rates and lead to an excellent discovery prospect for the near future. For a Higgs mass $m_h\simeq 125$~GeV and a pNGB
characteristic scale $f \lesssim 1$ TeV, we find that the DM mass is either $m_\eta \simeq 50-70$ GeV, with DM annihilations driven by the Higgs resonance,
or  in the range $100-500$~GeV, where the DM derivative interaction with the Higgs becomes dominant. In the former case the invisible Higgs decay to two DM particles could weaken the LHC Higgs signal.

 \end{abstract}
\newpage

%%%%%%%%%%%%%%%%%%%%%%%%%%%%%%%%%%%%%%%%%%%%%%%%%%%%%%%%%%%%%%%%%%%%%%%%%
\section{Motivation}\label{sec:motivation}
%%%%%%%%%%%%%%%%%%%%%%%%%%%%%%%%%%%%%%%%%%%%%%%%%%%%%%%%%%%%%%%%%%%%%%%%%

From a theoretical point of view light  scalar particles are unnatural, unless
a suitable structure protects their mass from large quantum corrections.
In the Standard Model (SM) there is a compelling case for a light scalar, the Higgs boson, 
and many efforts have been made to address the associated hierarchy problem.
The dark matter (DM) energy density of the Universe
could also be  accounted for by  a new light scalar.
The minimal realization, extensively studied in the literature (see e.g. Ref.~\cite{McDonald:1993ex}), 
consists in  adding to the SM a gauge singlet real scalar $\eta$, and  assuming that it is stable
due to a  parity  $\eta \rightarrow -\eta$. 
The model  is quite  predictive since it only depends on two extra
parameters: the singlet mass $m_\eta$ and its ``portal" coupling to the Higgs boson, $\lambda$. 
 This apparent simplicity, however, calls for an ultraviolet completion.
As we said, light scalars are unnatural in  quantum field theories, unless they are accompanied by  new 
ingredients,  such as for example supersymmetry,  and these deeply 
affect the dynamics  of  the DM.

Light scalars can also be natural if they are not elementary particles, rather composite states
emerging from a  strongly-coupled fundamental theory. 
Similarly as    pions in QCD,  light scalars can appear  as pseudo Nambu-Goldstone bosons (pNGBs)  arising from the  spontaneous breaking of the global symmetries of the  strong sector
around the $\TeV$ scale.  
A well-known example 
is the symmetry breaking pattern $SO(5)/SO(4)$, that generates  four pNGBs with the
quantum numbers of the SM Higgs doublet $H$, the weak gauge group being embedded in $SO(4)\simeq SU(2)_L\times SU(2)_R$~\cite{CH:ACP}.
The most  minimal extension to this  symmetry pattern  is the   $SO(6)/SO(5)$ model,  that contains
five pNGBs in the spectrum: the Higgs doublet $H$ and a gauge singlet $\eta$  \cite{Gripaios:2009pe}.
Interestingly, the pattern $SO(6)/SO(5)$ is
the minimal   example  with a known ultraviolet completion in terms of techni-quarks  \cite{Galloway:2010bp}.

The purpose  of this article is to study under which conditions
 $\eta$  is a suitable DM candidate. 
The properties of  $\eta$, being a  pNGB,  are  determined by the global symmetries of the strong
sector, and by the way these symmetries are  explicitly  broken by  the couplings of the SM fields  to the strong  sector.
We will  analyze the phenomenology of this composite DM particle  and its interplay with the composite Higgs doublet $H$. 

The DM couplings are substantially different from the non-composite singlet case.
This is mainly due to new non-renormalizable interactions
between $\eta$ and the SM fields, arising from operators of dimension-six 
suppressed by   $1/f^2$ where   $f\sim$ TeV is the decay constant  of the NGBs.
There are two types of these interactions: 
  (i) derivative couplings  between $\eta$ and $H$, fully determined by  
the $SO(6)/SO(5)$ structure,  
that scale as $p^2/f^2$ where $p$ is the relevant momentum in a given process,
and (ii)  direct couplings between $\eta$ and the SM fermions,
arising  from the explicit breaking of the global $SO(6)$ symmetry,
that scale as $m_\psi p/f^2$ 
where $m_\psi$ is the mass of the fermion. 
These   interactions 
make the  phenomenology of the composite scalar DM  substantially 
different from that of an elementary scalar (similar candidates of composite scalar DM have been also proposed, for instance, in the context of techni-colour theories \cite{Ryttov:2008xe} and 
gauge-Higgs unified models \cite{Hosotani:2009jk}; see also Ref.~\cite{baryonDM}).

In section \ref{sec:symm}
we present the effective lagrangian for the composite $\eta$, defined  
from  symmetry considerations  and naive dimensional analysis (NDA). 
In section \ref{Sec:RelicDensity} we compute the composite DM relic density.
In sections \ref{sec:LHC} and \ref{Sec:DirectDetection} we study respectively
the constraints coming from the Higgs searches at the LHC
and the DM direct detection experiments. We combine these results in section
\ref{sec:combined}, where we identify the regions of parameters 
that fulfill all the
phenomenological requirements. We conclude in section \ref{sec:conc}  and leave the
technical details of the composite models for  appendix \ref{sec:theory}, and 
the lengthy expressions for the relic density and direct detection cross sections for appendices
\ref{app:A} and \ref{app:B}, respectively.

%%%%%%%%%%%%%%%%%%%%%%%%%%%%%%%%%%%%%%%%%%%%%%%%%%%%%
\section{Light scalar dark matter from a composite sector
}\label{sec:symm}
%%%%%%%%%%%%%%%%%%%%%%%%%%%%%%%%%%%%%%%%%%%%%%%%%%%%%

We consider  theories  with a  light scalar  sector consisting only of a Higgs doublet $H$ and 
a singlet $\eta$ with parity 
\begin{equation}\label{z2new}
\eta\rightarrow -\eta~,
\end{equation}
that  makes $\eta$ stable.
  The lightness of these states is a consequence of a global symmetry.
The simplest realization consists of  having a new
strong sector  with
 a global symmetry breaking pattern  given by $O(6)\rightarrow O(5)$.
Five NGBs   emerge from this spontaneous breaking: a ${\bf 5}$ of $SO(5)$ which decomposes as
$\mathbf{4} \oplus \mathbf{1} \simeq (\mathbf{2},\mathbf{2}) \oplus
(\mathbf{1},\mathbf{1})$
under  the subgroup $SO(4) \simeq SU(2)_L \times SU(2)_R$. The $\mathbf{4}$  is
identified with   $H$, while the gauge singlet is $\eta$ (see appendix A for details).
Thinking of $SO(6)$ as the rotation group in a six-dimensional space, 
the symmetry under which
$\eta$ shifts  corresponds to a  rotation in the 5-6 plane, 
$SO(2)_{5-6}\equiv SO(2)_{\eta}\simeq U(1)_\eta$, while
\eq{z2new} corresponds to the six-dimensional parity  of $O(6)$, $P_{\eta}=\textrm{diag}(1,1,1,1,-1,1)$.

If it were for the strong sector alone, the pNGBs $H$ and $\eta$ would be massless and would only
interact derivatively.
However, the SM gauge bosons and fermions couple to the strong sector
breaking explicitly $O(6)$, generating non-derivative interactions between $H$, $\eta$ 
and the SM fields.
We will assume that $P_\eta$ is preserved by the SM couplings to the strong sector.
At the one-loop level these couplings induce a potential for $H$ and $\eta$, that is eventually
responsible for electroweak symmetry breaking (EWSB), with $v \equiv \sqrt{2} \langle H \rangle = 246$ GeV
and $\langle \eta\rangle =0$.

This scenario, apart from giving a solution to the hierarchy problem, since the Higgs is naturally a light state,
also provides  a good candidate for  DM,  the extra state $\eta$, that is also naturally light.
At energies below the strong scale, denoted by $m_\rho\sim$ few TeV,
the lagrangian for $\eta$ at the lowest order in a $\eta^2/f^2$ expansion, where $f$ is the pNGB decay constant, is given by
\begin{eqnarray}
\mathcal{L}_{\eta}&=& \frac{1}{2}(\partial_\mu \eta)^2-V(\eta,H)
+\frac{1}{2f^2}\left(\partial_{\mu}|H|^2+\frac{1}{2}\partial_{\mu}\eta^2\right)^2\label{eq:L1}\\
&+&\frac{\eta^2}{f^2} \left(c_t y_t \  \overline{q_L} \tilde H t_R+ c_b y_b \ \overline{q_L}H  b_R + h.c. \right) +
\cdots\, ,\label{eq:L2}
\end{eqnarray}
where
\begin{equation}
V(\eta,H)=\frac{1}{2}\mu_\eta^2 \eta^2+\lambda |H|^2\eta^2+\cdots ~.
\label{potential}
\end{equation}
For simplicity, only the interactions with the third family of SM quarks are shown, that are the ones that
play the most prominent role.  
Let us remark that these operators are not suppressed by the mass of  the heavy
composite states $m_\rho$  but  rather by  the  smaller scale 
$f \sim  m_\rho/g_\rho$  where 
$g_\rho$ is the inter-composite coupling, expected to satisfy $1\lesssim g_\rho\lesssim 4\pi$.

A few comments on  the lagrangian in Eqs.~(\ref{eq:L1})-(\ref{potential}) are in order.
The only interaction that preserves the Nambu-Goldstone shift symmetries is the derivative term in \eq{eq:L1} 
that, as shown in appendix A,  is fully determined by the  $SO(6)/SO(5)$ symmetry.
The  coefficients $c_{t,b}$, instead, depend crucially on how the fermions couple to the strong sector
and break explicitly the $SO(6)$ symmetry.
In general, we expect $c_{t}$ ($c_b$) to be  an  $O(1)$ complex number
whenever the  $U(1)_\eta$ symmetry is  broken
by the  $q_L$ and/or by the $t_R$ $(b_R)$  coupling to the strong sector;
otherwise they are zero. Both possibilities are realized in simple models, as described in appendix A.
Notice  that the interactions of \eq{eq:L2}
can also be induced from operators of the form
\begin{equation}\label{psidirectcoupling}
\overline{q_L} 
\gamma^{\mu}\partial_\mu q_L \frac{\eta^2}{f^2} ~,
\end{equation}
and similarly for $t_R$ and $b_R$.
Indeed, using the equations of motion for the fermion
we can rewrite the above  operator as those in \eq{eq:L2}.

Let us briefly discuss the  potential $V(\eta, H)$.
It  can only arise from loop effects
involving the SM fields.
Since the SM gauge interactions preserve the $U(1)_\eta$ symmetry,
$\mu_\eta$ and $\lambda$  can only be generated from fermion interactions
that break $U(1)_\eta$.
Since $\lambda$ is further protected by the symmetry  under which $H$ shifts,  we  generically expect
\begin{equation}\label{expectTheo}
\lambda \lesssim \mu_\eta^2/f^2 \, .
\end{equation}
Then, the mass of $\eta$ is given by $m_\eta^2=\mu_\eta^2+\lambda v^2\simeq \mu_\eta^2$, since 
electroweak precision measurements require \cite{Giudice:2007fh} 
\begin{equation}\label{defxi} 
 \xi\equiv \frac{v^2}{f^2} \ll 1\, .
 \end{equation}
It is easy to construct models where the $U(1)_\eta$ symmetry is either broken by the top
or the bottom coupling to the strong sector,  giving a one-loop mass for $\eta$
of the order of 500 or 50 GeV respectively,
as shown in appendix A. We also notice
that  in the potential  we have ignored  terms beyond the quadratic order in the  $\eta$ field,  since they do not 
play any important role in our DM analysis.

Apart from  \eq{eq:L1} and \eq{eq:L2},
there is also  the possibility to have at  order $\eta^2/f^2$ the  interaction terms
\begin{equation}\label{DirectGauge}
\frac{\eta^2}{f^2} \sum_{F=B,W,G} \left(c_{F\eta} F_{\mu\nu}F^{\mu\nu} + \tilde{c}_{F\eta} F_{\mu\nu} \tilde{F}^{\mu\nu} \right) \, .
\end{equation}
These couplings  do not respect the $U(1)_\eta$ symmetry
and therefore can only be induced by loops
involving  heavy composite states  that   see the $U(1)_\eta$-breaking 
through mixing with the SM states. 
We estimate them to be $c_{F\eta},\tilde c_{F\eta}\sim  (g_F/g_\rho)^2 (m_\eta/m_\rho)^2$, smaller than
contributions coming from SM loops.
We will  neglect them from now on.

In summary, the interactions between our DM candidate $\eta$ and the SM particles are controlled by 
four parameters: $c_b$, $c_t$,  $f$ and  $\lambda$. 
For our DM analysis below  we shall consider the following values for these parameters.
Since the  bottom quark plays a main role in the computation of the DM relic density,  
 in order to highlight the effects of compositeness
we will assume that its couplings to the strong sector break $U(1)_\eta$, so that $c_b\ne0$. 
More specifically, motivated by the models described in appendix A, we will consider the range
\be
c_b =\frac 12 + a + i b~, ~~~~ a,b \in [0,1]~,
\label{param}\ee
 where small (large) values of $a$ and $b$ correspond to small (large) couplings of $b_{L,R}$ to the strong sector.
For the top coupling $c_t$ we will consider two cases:
\begin{itemize}
\item Case 1: $U(1)_\eta$ is broken by the top and we fix for definiteness $c_t=1/2$ as in minimal models
(see appendix A). The top loops make $\eta$ generically heavier than the Higgs.
\item Case 2: The top interactions do not break $U(1)_\eta$. We then have $c_t=0$ and $m_\eta$ does not receive contributions from top loops.
In this scenario the DM particle $\eta$ can be naturally lighter than the Higgs.
\end{itemize}
For the decay constant $f$ we  will take  $f=500$  GeV and $f=1$  TeV.
Recall that although composite Higgs models  generically predict  
$f\sim v\simeq 246$ GeV,
electroweak precision  measurements require \eq{defxi} to hold.
Finally,   we will leave   $\lambda$ and $m_\eta$ as  free parameters to be determined 
by the requirement of $\eta$ to be a realistic  DM candidate.

The couplings of $\eta$ to the 1st and 2nd family  quarks will  be  important only when
considering direct DM detection.
If not otherwise specified,  we will assume for simplicity family independent coefficients,
$c_u=c_c=c_t$ and $c_d=c_s=c_b$.
This choice is partly motivated by the bounds on flavour violation \cite{Gripaios:2009pe}.

To recover the predictions of the renormalizable model \cite{McDonald:1993ex},
that we will refer to as the `non-composite case', we can take  the limit $f\rightarrow \infty$.
In this case only  $\lambda$  controls the interactions of $\eta$ with the SM fields.

%%%%%%%%%%%%%%%%%%%%%%%%%%%%%%%%%%%%%%%%%%%%%%%%%%%%%
%%%%%%%%%%%%%%%%%%%%%%%%%%%%%%%%%%%%%
\section{Relic density of the composite dark matter}
\label{Sec:RelicDensity}
%%%%%%%%%%%%%%%%%%%%%%%%%%%%%%%%%%%%%

In the standard cosmological framework,  DM was  kept in thermal and chemical equilibrium as a consequence of its interactions with the SM particles. As the Universe expanded and cooled, the number density of  DM particles decreased until they could not annihilate anymore, freezing out by the primordial thermodynamical equilibrium. Thenceforth their number density  remained constant, fixed to 
the value observed today.
This picture is described by a Boltzmann equation  that,  under certain assumptions, has the useful approximated solution~\cite{Gondolo:1990dk}
\begin{equation}\label{eq:GoldenRule}
\Omega_{\eta}h^2\simeq \frac{3\cdot 10^{-27}{\rm cm}^{3}{\rm s}^{-1}}{
\langle \sigma v_{rel} \rangle}~,
\end{equation}
where $\Omega_{\eta}=\rho_{\eta}/\rho_c$ is the ratio between the energy density of  DM and the critical energy density of the Universe,
$h=H_{0}/(100\,{\rm km}\,{\rm s}^{-1}{\rm Mpc}^{-1})$  is the reduced value of the present Hubble parameter,  and $\langle \sigma v_{rel} \rangle$
is the thermal-average of the total annihilation cross section of  DM particles,  times their relative velocity. The present experimental value is $\Omega_{DM}h^2=0.1126\pm 0.0036$ \cite{Komatsu:2010fb}.\footnote{
An alternative mechanism to generate the DM relic density is freeze-in, that is realized when the DM interactions with the SM are so weak that 
the DM species was never in equilibrium. In this case SM particles can slowly annihilate (or decay) into DM generating the required relic density.
For the case of a scalar singlet $\eta$ coupled to the Higgs portal, freeze-in is possible both for $m_\eta$ of the order of the electroweak scale \cite{Yaguna}
and for $m_\eta \lesssim 1$ GeV \cite{Frigerio}, as long as $\lambda \lesssim 10^{-10}$. Here we will consider only composite models that induce a larger value for $\lambda$ as well as a large DM-Higgs derivative interaction. Therefore we will focus on the freeze-out scenario.}

Although the singlet pNGB $\eta$ interacts strongly with the heavy resonances of the composite sector, its interactions with the SM,
including the composite Higgs, are weak, as shown in Eqs.~(\ref{eq:L1})-(\ref{potential}); 
as a consequence, its annihilation cross section into fermions, gauge bosons 
and the Higgs falls into the ballpark suggested by Eq. (\ref{eq:GoldenRule}).
In appendix \ref{app:A} we display all the relevant annihilation cross-sections that enter in the Boltzmann equation, both in the non-composite and in the composite case. The Feynman rules needed
in the computation are collected in Table~\ref{fig:FRulesCDM}.

The numerical solution of the Boltzmann equation is shown in Fig.~\ref{fig:BoltzmannRelic} as a function of $m_\eta$,
for two representative values of $\lambda$. 
To understand the behavior of the relic density with $m_\eta$, 
one should keep in mind that each annihilation channel $\eta\eta \leftrightarrow X\bar{X}$ opens only for $\eta$ masses above the threshold value
$ m_X \sqrt{1-v_{rel}^2/4}$, with $v_{rel}\sim 2/3$ at freeze-out.
In order to analyze the effect of the composite interactions,
let us recall that in the non-composite case the annihilation cross section is entirely determined by the value of $\lambda$ 
(see Eqs. (\ref{eq:AnnihilationSSBottom}-\ref{eq:AnnihilationSSH})).
In the composite case the situation is substantially different.
One can identify four interesting mass regions:

\begin{figure}[!t!]
  \begin{minipage}{0.4\textwidth}
   \centering
   \includegraphics[scale=0.65]{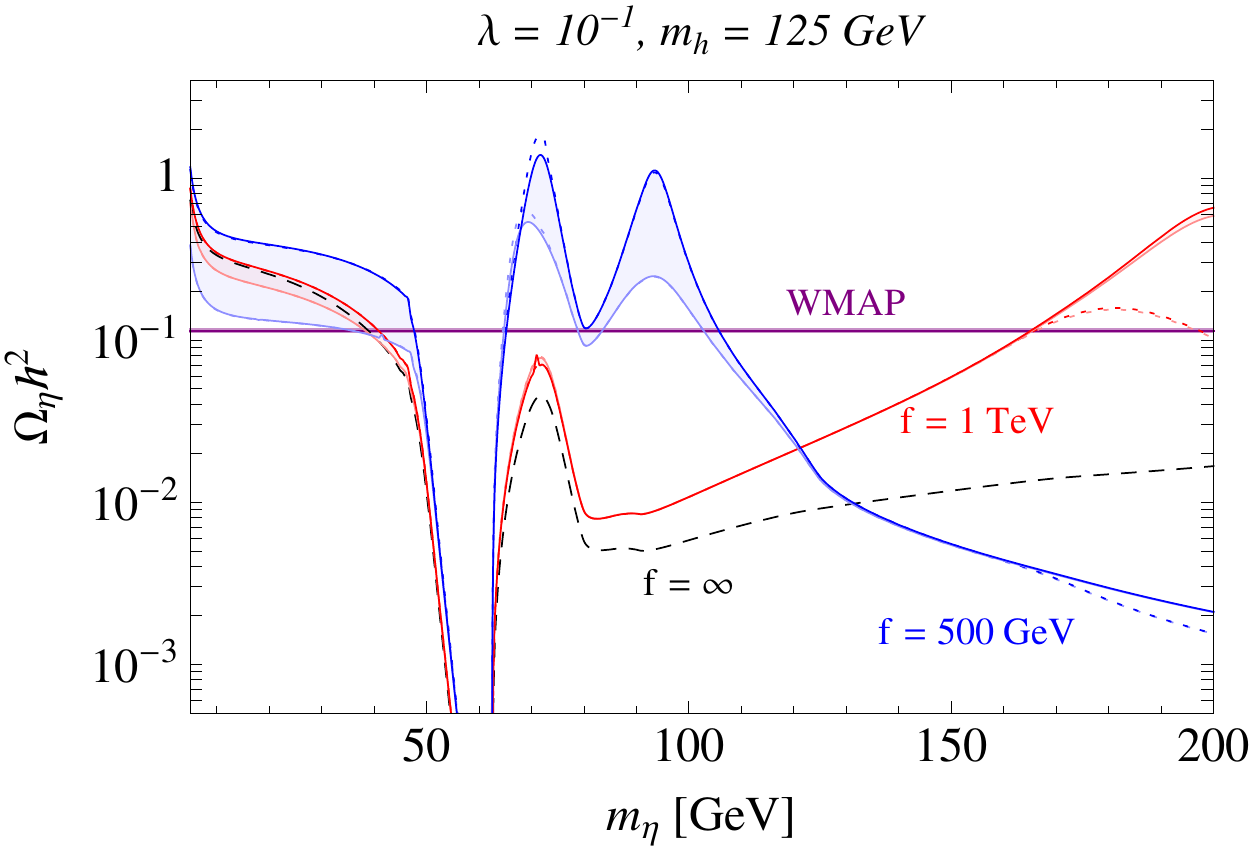}
    \end{minipage}\hspace{1.5 cm}
   \begin{minipage}{0.4\textwidth}
    \centering
    \includegraphics[scale=0.65]{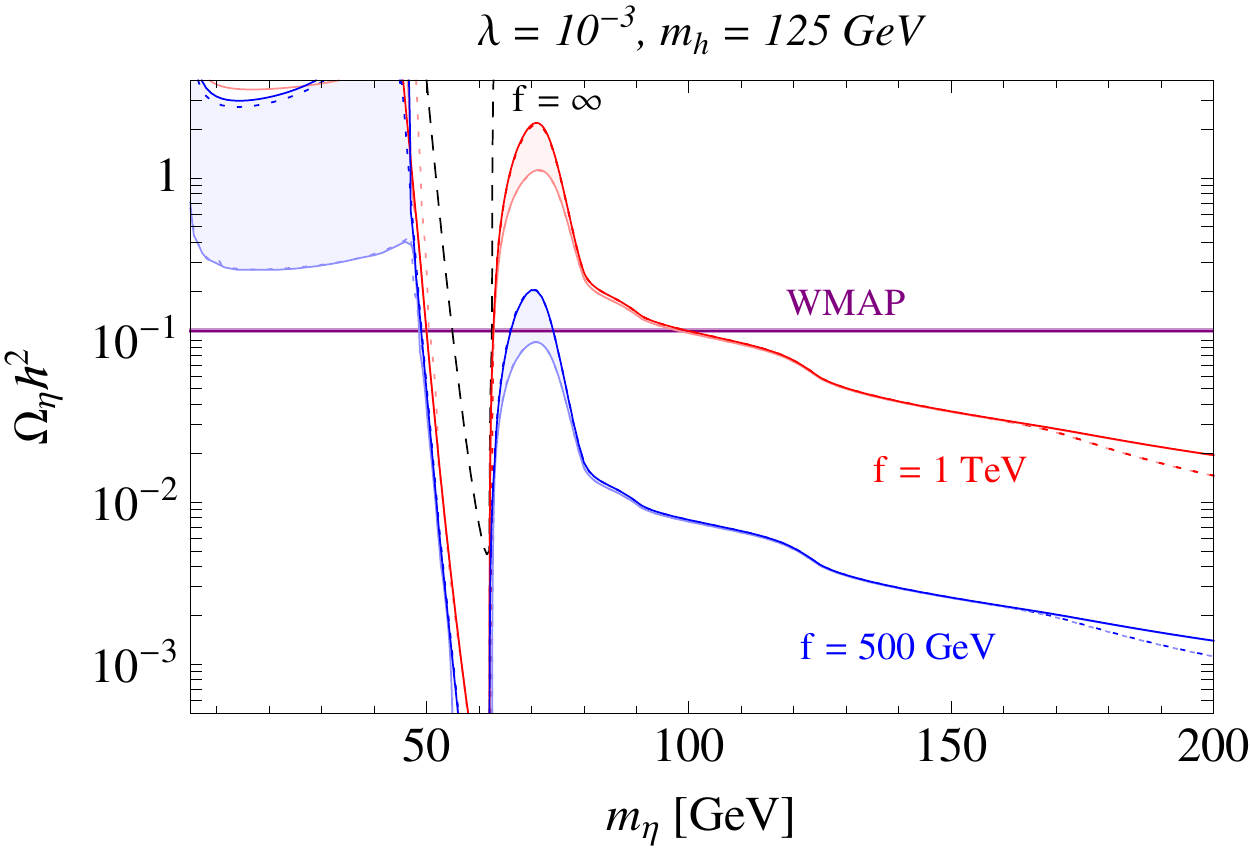}
    \end{minipage}
 \caption{\emph{Relic density of the scalar singlet DM as a function of its mass $m_{\eta}$.
We take $m_h=125$ GeV and $\lambda=10^{-1}$ ($10^{-3}$) in the left (right) panel. 
The dashed curves correspond to the non-composite case, the red (blue) curves to the composite case with $f=1$ TeV 
($500$ GeV).
The red and blue bands describe the variation of the DM$-$bottom coupling $c_b$ in \eq{param}, 
between $a=b=0$ (dark lines) and $a=b=1$ (light lines).
Above the threshold for annihilation into $t\bar{t}$, the result depends also on the value of the DM$-$top coupling:
the dotted and solid lines correspond to $c_t=1/2$ (Case 1) and $c_t=0$ (Case 2), respectively.}} 
 \label{fig:BoltzmannRelic}
\end{figure}

\begin{enumerate}

\item \underline{The low-mass region, $ m_\eta \lesssim 50 \GeV$.} The annihilation cross-section is dominated by the $b\bar{b}$ channel, 
that is enhanced by the direct DM$-$bottom coupling proportional to $c_b$, shown in \eq{eq:L2}. 
As a consequence, one can reproduce the DM relic density even for values of $\lambda$ smaller than in 
the non-composite case.

\item \underline{The resonant region, $m_{\eta}\simeq m_{h}/2$.} The presence of the Higgs resonance enhances 
the annihilation cross section  by several orders of magnitude. In the non-composite case this is the only possibility 
to fit the DM relic density when $\lambda$
is relatively small, however for too small values ($\lambda \lesssim 10^{-4}$ at $m_h=125$ GeV) even this enhancement is not sufficient
and the model is excluded.
On the contrary, in the composite case one can reproduce the DM relic density even for vanishing $\lambda$, 
because of the extra derivative contribution to the $\eta$-$\eta$-$h$ vertex, coming from \eq{eq:L1}.

\item \underline{The cancellation region, $m_\eta^2 \sim \lambda f^2 /2$.} 
A cancellation can occur between the derivative and the $\lambda$  contributions  to the $\eta$-$\eta$-$h$ vertex
(see \eq{eq:L1} and  \eq{potential}, respectively), that may suppress the annihilation with s-channel  Higgs-exchange when this is too large. 
The cancellation condition reads
\begin{equation}\label{eq:Cancellazioni}
s= 2  \lambda f^2 (1-\xi)~,
\end{equation}
where $s=4m_{\eta}^2/(1-v_{rel}^2/4)$. 
For example, for $\lambda=10^{-1}$ and $f=500$ GeV, taking into account the freeze-out temperature one finds 
that the effect of the cancellation is maximal for $70$ GeV $\lesssim m_{\eta}\lesssim 100$ GeV,
as one can see in the left panel of Fig.\ref{fig:BoltzmannRelic}.
In the non-composite case ($f\rightarrow \infty$), when $\lambda$ is sufficiently large
the relic density is too suppressed for all values of $m_\eta$ above the resonant region. In the case of compositeness, instead,
the cancellation enhances the value of $\Omega_\eta$ and can make it compatible with the DM relic density.
For larger $f$, the cancellation region moves out to larger values of $m_\eta$, where the $t\bar{t}$ channel opens.
If the top has no direct coupling to $\eta$ 
(solid lines in Fig.~\ref{fig:BoltzmannRelic}),
the cancellation remains effective. 
If instead the top couples directly to DM (dotted lines), 
this reduces the relic density already for values of $m_\eta$ in the cancellation region.

\item \underline{The high-mass region.}
As the DM mass increases,   $m^2_\eta\gg m_h^2/4$ and $m^2_\eta\gg \lambda f^2 /2$,
 the $\eta$-$\eta$-$h$ vertex
is more and more dominated by the derivative coupling of Eq. (\ref{eq:L1}), whose strength is uniquely fixed by $f$.
As a consequence, the annihilation rate through the Higgs portal becomes larger 
(typically too large for $f= 500 \GeV$).
Still, for $f\lesssim 1\TeV$ the annihilation rate is of the correct order of magnitude, as can be seen in the right panel of Fig. \ref{fig:BoltzmannRelic}. The proper relic density can be reproduced even for a vanishing $\lambda$. 

\end{enumerate}

The analysis of the $\eta$ relic density will be completed
in section \ref{sec:combined}, where we will show the contours for $\Omega_\eta = \Omega_{DM}$ in the plane $(m_\eta , \lambda)$.
Before that, we need to discuss the
constraints coming from LHC Higgs searches (section \ref{sec:LHC}), and from DM direct detection experiments (section \ref{Sec:DirectDetection}).

%%%%%%%%%%%%%%%%%%%%%%%%%%%%%%%%%%%%%%%%%%%%%%%%%%%%%
\section{Constraints from Higgs searches at the LHC}\label{sec:LHC}
%%%%%%%%%%%%%%%%%%%%%%%%%%%%%%%%%%%%%%%%%%%%%%%%%%%%%

The SM Higgs boson mass was restricted to be heavier than 114 GeV by LEP measurements.
This limit applies also to a Higgs boson that decays invisibly, since even in this case the bound from associated production with a $Z$ boson 
holds \cite{:2001xz}. 
The LHC 95\% C.L. upper bound in the low mass region is presently 127 (129) GeV from the CMS (ATLAS) experiment, assuming a
SM-like Higgs. Here we will not discuss the very high mass region ($m_h\gtrsim 600$ GeV) that is still allowed.\footnote{
We have checked that, 
even for a very heavy Higgs boson, with $m_h\gtrsim 600\GeV$, the composite singlet can still provide a good DM candidate.}
In the low mass region there is also a hint for a Higgs with $m_h\simeq 125\GeV$, coming mostly from the decay channel
$h\rightarrow \gamma\gamma$ \cite{ATLAS}.

In the singlet extension of the SM, invisible Higgs decays are allowed as long as $m_h > 2m_\eta$.
The observation of the Higgs boson would imply an upper bound on the invisible decay width,
$\Gamma_{inv}\equiv\Gamma(h\rightarrow\eta\eta)$. Vice versa, a signal suppression could be explained by a sizable $\Gamma_{inv}$.
In our scenario we find
\begin{equation}\label{eq:IDW}
\Gamma_{inv} =
\frac{v^2}{32\pi m_h}
\sqrt{1-\frac{4m_{\eta}^2}{m_h^2}}
\left(\frac{m_h^2}{v^2}\frac{\xi}{\sqrt{1-\xi}}-2\lambda \sqrt{1-\xi}\right)^2
\vartheta(m_h-2m_{\eta})~.
\end{equation}
In addition to the invisible decay channel,  compositeness implies further deviations from the SM predictions, 
because the Higgs couplings to gauge bosons and fermions are modified at order $\xi$ \cite{Giudice:2007fh}, as we will describe below.
These modifications affect both the Higgs production and the relative branching ratios for the Higgs decays into visible channels.
Reduced (enhanced) couplings would weaken (strengthen) the Higgs signal  at the LHC.

Let us discuss first the non-composite scenario, when $\xi$ vanishes. In this case the LHC exclusion limits on the Higgs mass 
are conveniently described by the ratio
\be
\mu \equiv \frac{\sigma(pp\rightarrow h \rightarrow SM)}{\sigma_{SM}(pp\rightarrow h \rightarrow SM)}
\equiv \frac{\sigma(pp\rightarrow h)}{\sigma_{SM}(pp\rightarrow h)} 
BR(h\rightarrow SM)  = \frac{\Gamma_{SM}}{\Gamma_{SM}+\Gamma_{inv}} ~,
\label{mu}\ee 
where $\Gamma_{SM}$ is the total Higgs width in the SM, and 
$\Gamma_{inv}$ is given by \eq{eq:IDW} in the limit $\xi\rightarrow 0$.
Note that the Higgs production cross section is unchanged w.r.t. the SM,
however the Higgs visible branching ratio becomes smaller than one, as long as $m_h>2m_\eta$.
Actually for $|\lambda|\gtrsim 0.1$ the channel $h\to \eta\eta$ dominates.

In the case of composite $h$ and $\eta$, there are several important differences. First, there is another $h-\eta-\eta$ coupling besides $\lambda$,
that arises from the derivative interaction 
in \eq{eq:L1}. 
As a consequence, 
the two contributions to the decay amplitude in \eq{eq:IDW}  can cancel each other for $\lambda>0$,
or  add up for $\lambda<0$. In the former case the Higgs signal at the LHC is maximal for
$\lambda \simeq m_h^2/2f^2$, rather than for $\lambda=0$.

Second, as we mentioned, there are order $\xi$ modifications of the Higgs couplings, that are specific to the composite nature of $h$ 
independently from the existence of a light singlet $\eta$.
For concreteness,  we will center here  on the  $SO(6)/SO(5)$ models described in appendix A, assuming Case 2 for the top quark.
The relevant Higgs couplings are modified as follows:
\bea
g_{hVV} &=& g_{hVV}^{SM} \sqrt{1-\xi}~~~~{\rm for}~V=W,Z,\label{hVV}\\
g_{ht\bar{t}} &=& g_{ht\bar{t}}^{SM}\sqrt{1-\xi}~,\label{htt}\\
g_{hb\bar{b}} &=& g_{hb\bar{b}}^{SM}\frac{1-2\xi}{\sqrt{1-\xi}}~.
\label{hbb}
\eea
The Higgs coupling to gluons, which is crucial for the total production cross section, is mainly generated by top loops involving
the coupling $g_{ht\bar t}$ linearly. Therefore we have\footnote{
One should remark that, in composite Higgs model, the suppression of the top quark contribution to $g_{hgg}$ could be compensated by the contributions of heavier states, such as vector-like fermions that accompany the top, 
or other resonances of the strongly-interacting sector. However,
in most cases it was found that no enhancement of the Higgs production occurs, but exceptions are possible \cite{Falkowski:2007hz}.}
\be
g_{hgg} \simeq g_{hgg}^{SM} \frac{g_{ht\bar{t}}}{g_{ht\bar{t}}^{SM}} =  g_{hgg}^{SM}\sqrt{1- \xi} ~.
\label{hgg}\ee
Similarly to the case of gluons, 
the Higgs coupling to photons is dominated by a top quark loop and a $W$-loop, that involve linearly $g_{ht\bar{t}}$ and $g_{hWW}$ respectively.
Since both couplings are corrected in the same way (see Eqs.~(\ref{hVV}) and (\ref{htt})), one finds
\be
g_{h\gamma\gamma}\simeq g_{h\gamma\gamma}^{SM}\sqrt{1-\xi}~.
\label{hpp}\ee

In order to compare these effects of compositeness with the LHC results, 
one should notice that the most sensitive channels in the present analyses by ATLAS and CMS are
$h\rightarrow \gamma\gamma,WW,ZZ$. Therefore, what is actually measured is not the total Higgs signal strength $\mu$, rather the signal strength
in the `gauge' decay channels only. As a consequence, the relevant quantity is
\begin{align}
\mu_{VV} &\equiv \frac{\sigma(pp\rightarrow h \rightarrow VV)}
{\sigma_{SM}(pp\rightarrow h \rightarrow VV)} 
\equiv  \frac{\sigma(pp\rightarrow h)}{\sigma_{SM}(pp\rightarrow h)} 
\frac{\Gamma(h\rightarrow VV)}{\Gamma_{SM}(h\rightarrow VV)} \frac{\Gamma_{SM}}
{\Gamma_{SM}^{comp}+\Gamma_{inv}} \nonumber\\
&\simeq (1-\xi) (1-\xi)  
\frac{\Gamma^{\psi\bar{\psi}}_{SM}+\Gamma^{VV}_{SM}}
{\Gamma^{\psi\bar{\psi}}_{SM}(1-2\xi)^2/(1-\xi)+\Gamma^{VV}_{SM}(1-\xi) + \Gamma_{inv}}~,
\label{muvv}\end{align}
where $V=\gamma,Z,W$, and we assumed that the Higgs production is dominated by the gluon fusion channel,
controlled by \eq{hgg}.
Here $\Gamma_{SM}^{comp}$ is the total visible width in the composite scenario, while
$\Gamma^{\psi\bar{\psi}}_{SM}$ and $\Gamma^{VV}_{SM}$ are the total SM widths into fermions and gauge bosons, 
respectively. 
Note that, in the non-composite case, \eq{muvv} reduces to \eq{mu}.

\begin{figure}[!tb!]
   \centering
  \includegraphics[scale=0.64]{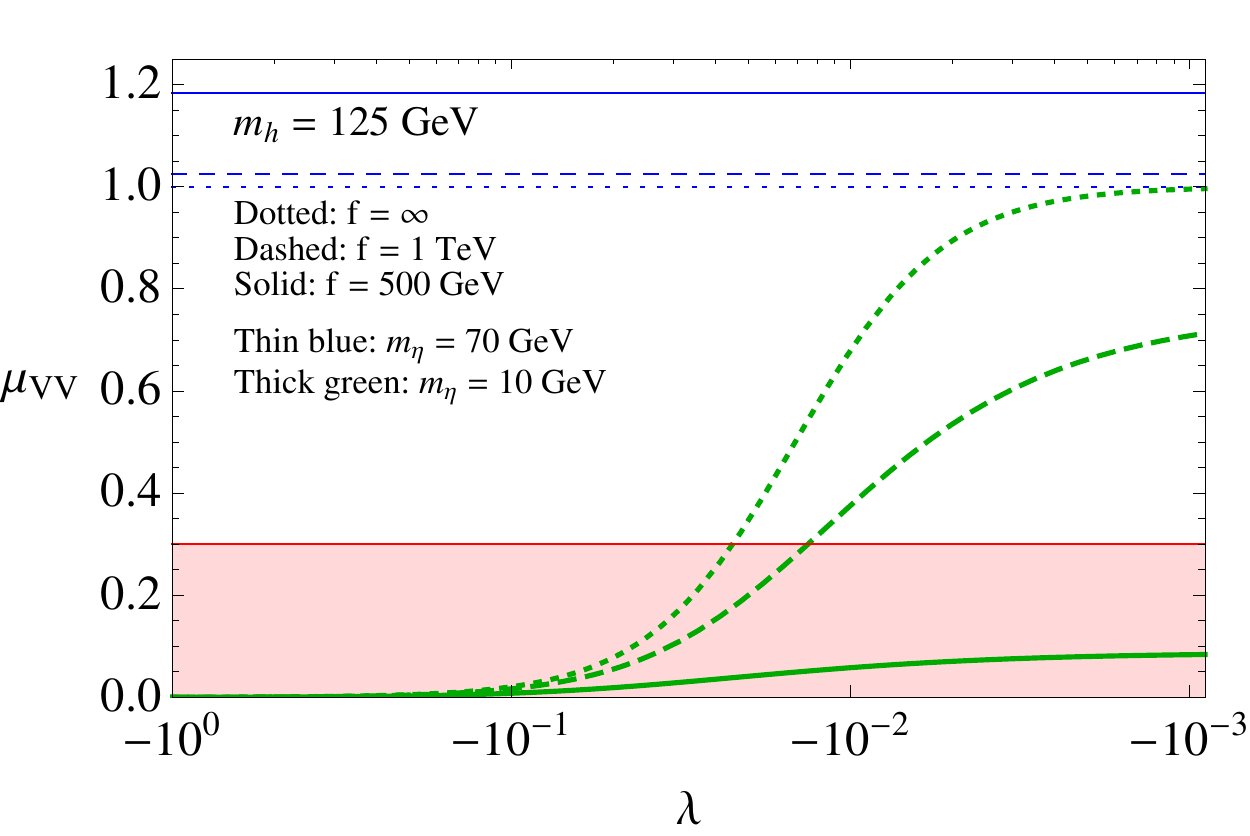}   
    \includegraphics[scale=0.64]{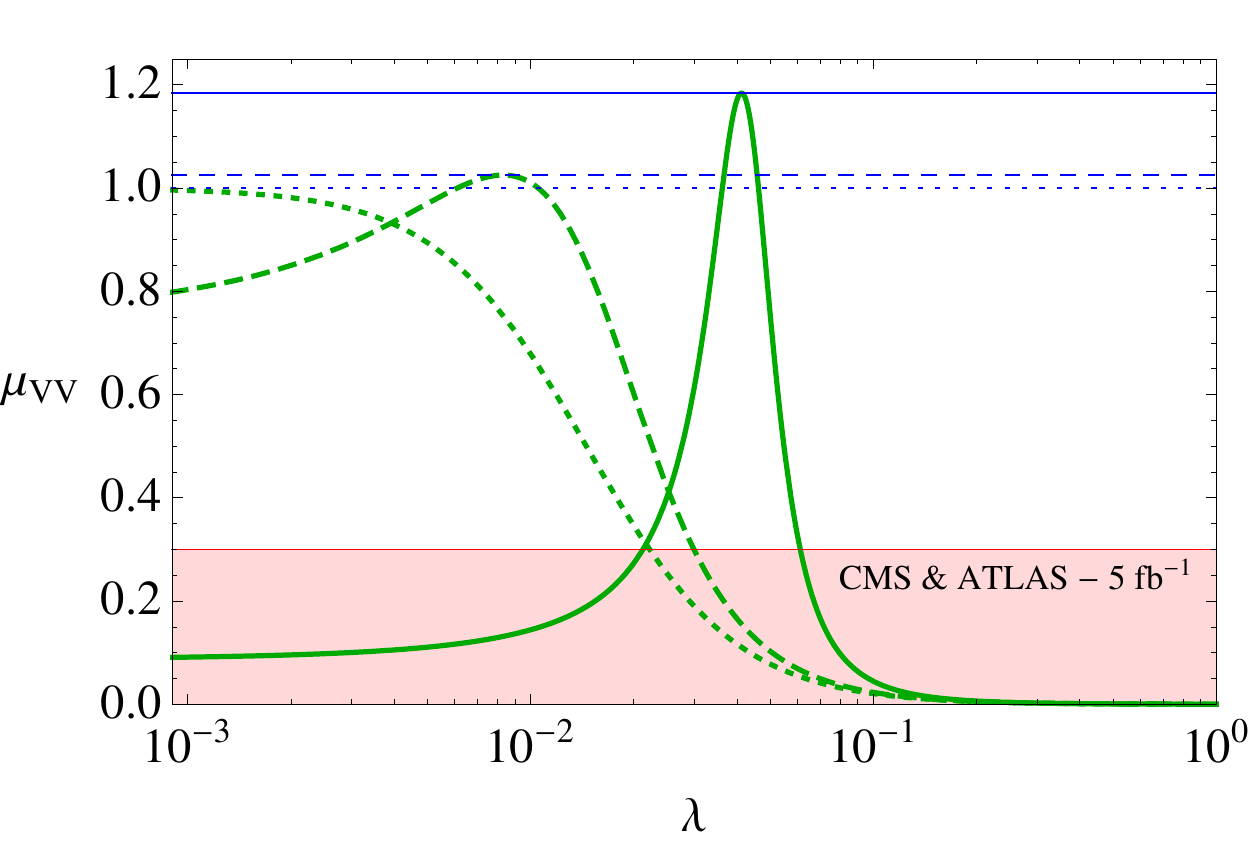}
  \caption{\emph{{The strength $\mu_{VV}$ of the Higgs signal in the gauge channels, defined in \eq{muvv},
as a function of the $h-\eta$ quartic coupling $\lambda$ (negative in the left panel, positive in the right panel).
We chose $m_h=125$ GeV and two values for the scalar singlet mass $m_\eta$, one larger than $m_h/2$ and the other smaller.
The dotted curves correspond to the non-composite case. The dashed (solid) curves correspond to compositeness with
 $f=1$ TeV (500 GeV). 
 We took into account the order $\xi$ corrections to the Higgs couplings to vector bosons and fermions,
 see Eqs.~(\ref{hVV})-(\ref{hpp}).
The shaded region is disfavoured by the LHC Higgs searches (we roughly extracted the 95\% C.L.
lower bound on $\mu_{VV}$ from Ref.~\cite{ATLAS}).
}}}
 \label{fig:lam125}
\end{figure}

\begin{figure}[!tb!]
   \centering
   \includegraphics[scale=0.64]{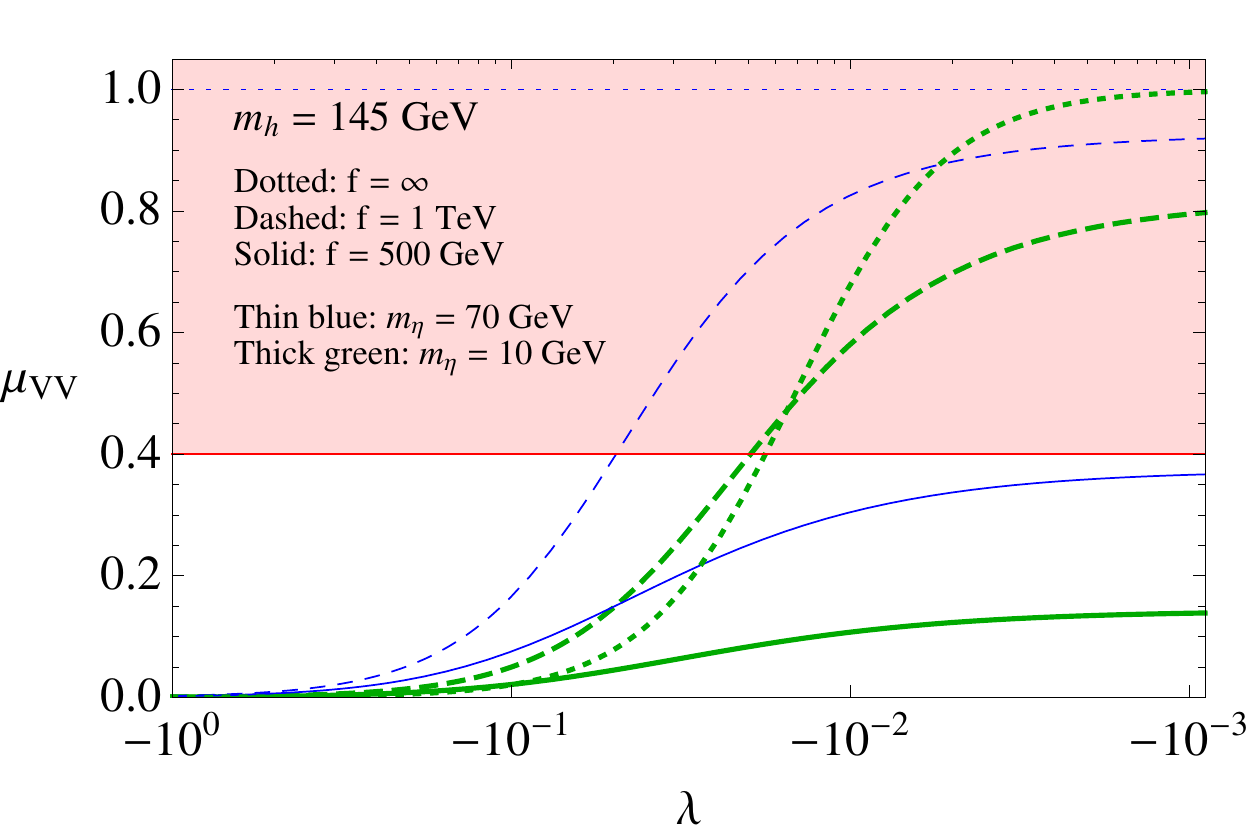}   
   \includegraphics[scale=0.64]{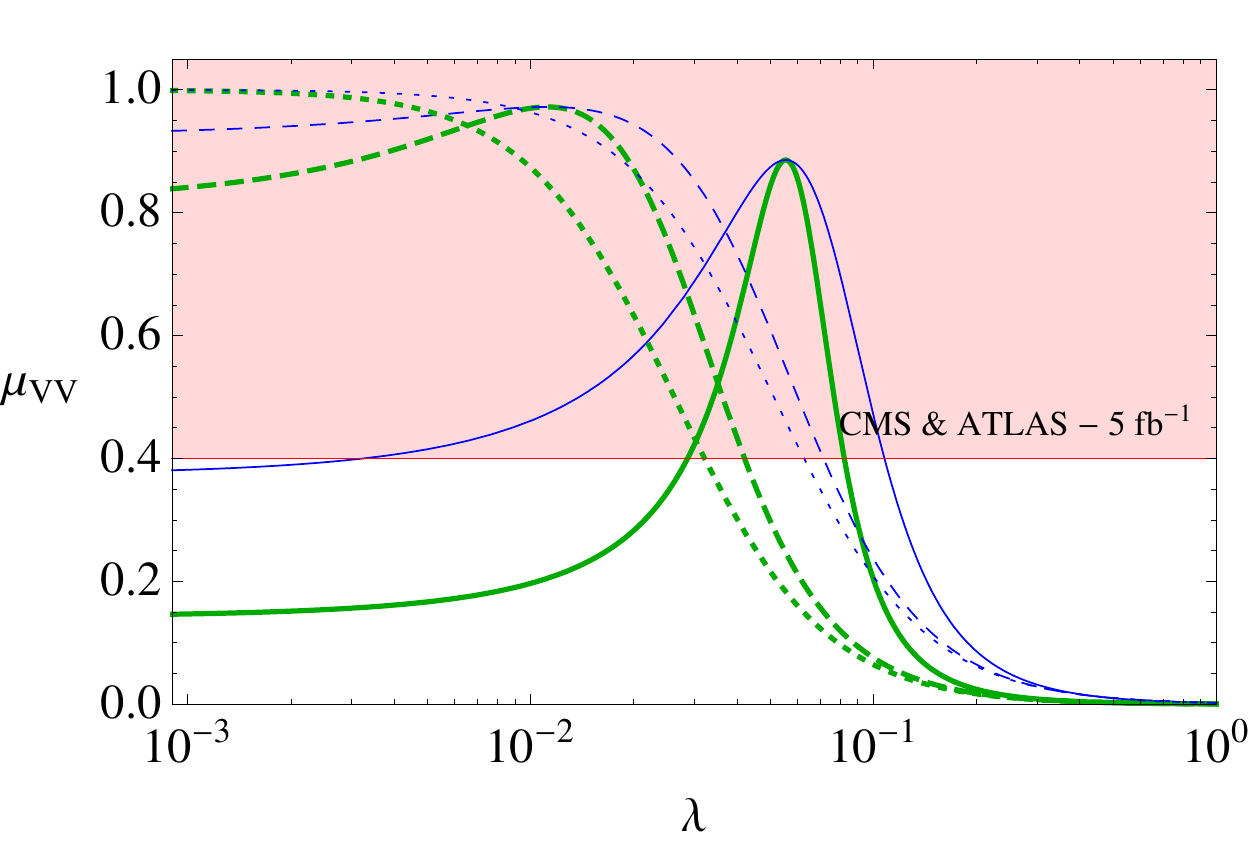}   
   \begin{minipage}{0.4\textwidth}
    \centering
       \end{minipage}
 \caption{\emph{
 The same as in Fig.~\ref{fig:lam125}, except for $m_h=145$ GeV. For this mass ATLAS and CMS put an upper bound on $\mu_{VV}$.}}
 \label{fig:lam145}
\end{figure}

In Figs.~\ref{fig:lam125},\ref{fig:lam145} we display our prediction for $\mu_{VV}$ as a function of $\lambda$.
Since the gauge channels are the most sensitive at the LHC, the experimental constraints on $\mu$ \cite{ATLAS}
are more appropriately interpreted as constraints on $\mu_{VV}$. The red shaded regions in Figs.~\ref{fig:lam125},\ref{fig:lam145} 
are disfavoured at $95\%$ C.L..
Let us discuss the `light' and 'heavy' Higgs scenarios in turn:\\

\noindent
\emph{$\bullet$ SM-like Higgs with $m_h\lesssim 130\GeV$} (Fig.~\ref{fig:lam125})

If the signal around $m_h=125$ GeV were confirmed, the ratio $\mu_{VV}$  should lie relatively close to one.
More precisely, the CMS (ATLAS) collaboration finds the largest excess at 
$m_h=124(126)$ GeV, with  $\mu_{VV} = 0.6 - 1.2~(0.6 - 1.3)$ at $1\sigma$ \cite{ATLAS}.
Given this large statistical uncertainty, in Fig.~\ref{fig:lam125} we roughly estimated the 
$95\%$ C.L. lower bound as $\mu \gtrsim 0.3$.

In the non-composite case, for $m_\eta>m_h/2$  one has $\mu_{VV}=1$, 
while for $m_\eta < m_h/2$, in order to avoid a too large $\Gamma_{inv},$ one is forced to take 
sufficiently small values of $\lambda$.

In the composite case instead, when $m_\eta>m_h/2$ we find that $\mu_{VV}$ can be larger than one.
Indeed \eq{muvv} shows that, even though the Higgs production is suppressed by a factor $(1-\xi)$, the gauge decay channels receive a $(1-\xi)^2/(1-2\xi)^2$ enhancement
relatively to the fermion channels (which dominate the branching ratio up to $m_h\simeq 135$ GeV). As a consequence, one finds 
that $\mu_{VV}$ can be as large as $1.2$  for $f$ as small as $500$ GeV.
When $m_\eta<m_h/2$, the signal is suppressed by $\Gamma_{inv}$.
However  larger 
values of $\lambda$ are allowed w.r.t. the non-composite case, as shown in the right panel of Fig.~\ref{fig:lam125}, 
because of the cancellation between the two terms in brackets in \eq{eq:IDW}.\\

\noindent
\emph{$\bullet$ Invisible Higgs with $m_h\gtrsim 130\GeV$} (Fig.~\ref{fig:lam145})

In this region the LHC searches exclude a SM-like Higgs, by requiring the ratio $\mu_{VV}$ to be smaller than one. In our scenario,
this can be accounted for by the combined effect of invisible decays and reduced SM couplings, as illustrated in
Fig.~\ref{fig:lam145}, where we fix $m_h=145$ GeV
(for this value of the Higgs mass the most sensitive channels are $h\rightarrow WW\rightarrow 2l2\nu$ and $h\rightarrow ZZ\rightarrow 4l$).
In this case the $95\%$ C.L. upper bound from ATLAS and CMS is roughly $\mu_{VV}\lesssim 0.4$.
There are wide regions of parameter space that survive the LHC constraint, as shown by the curves lying in 
the unshaded region in Fig.~\ref{fig:lam145}.

%%%%%%%%%%%%%%%%%%%%%%%%%%%%%%%%%%%%%
\section{Constraints from dark matter searches}
\label{Sec:DirectDetection}
%%%%%%%%%%%%%%%%%%%%%%%%%%%%%%%%%%%%%

The direct detection of DM is based on the observation of the elastic scattering between non-relativistic DM particles from our galaxy halo and cryogenic nuclei in targets, that produces a nuclear recoil. 
In our scenario the interaction of the composite DM $\eta$ with nucleons 
arises through the direct couplings $c_q$ $(q=u,d,...)$ between DM and quarks, given in  Eq.~(\ref{eq:L2}), and through the  t-channel exchange  of a Higgs boson,
which in turn couples with quarks; in both cases heavy quark loops also induce a coupling to the gluons in the nucleons.
These processes give rise to a spin-independent elastic cross section $\sigma_{\rm SI}$, potentially within the reach of present and future experiments. 
Note that ${\rm Im}(c_{q})$ contributes only to the spin-dependent cross-section $\sigma_{SD}$, which is always much smaller than $\sigma_{\rm SI}$
(the latter being enhanced by the coherent interactions on protons and neutrons in the large target nucleus).

\begin{figure}[t!]
 \begin{minipage}{0.40\textwidth}
  \centering
  \includegraphics[scale=0.65]{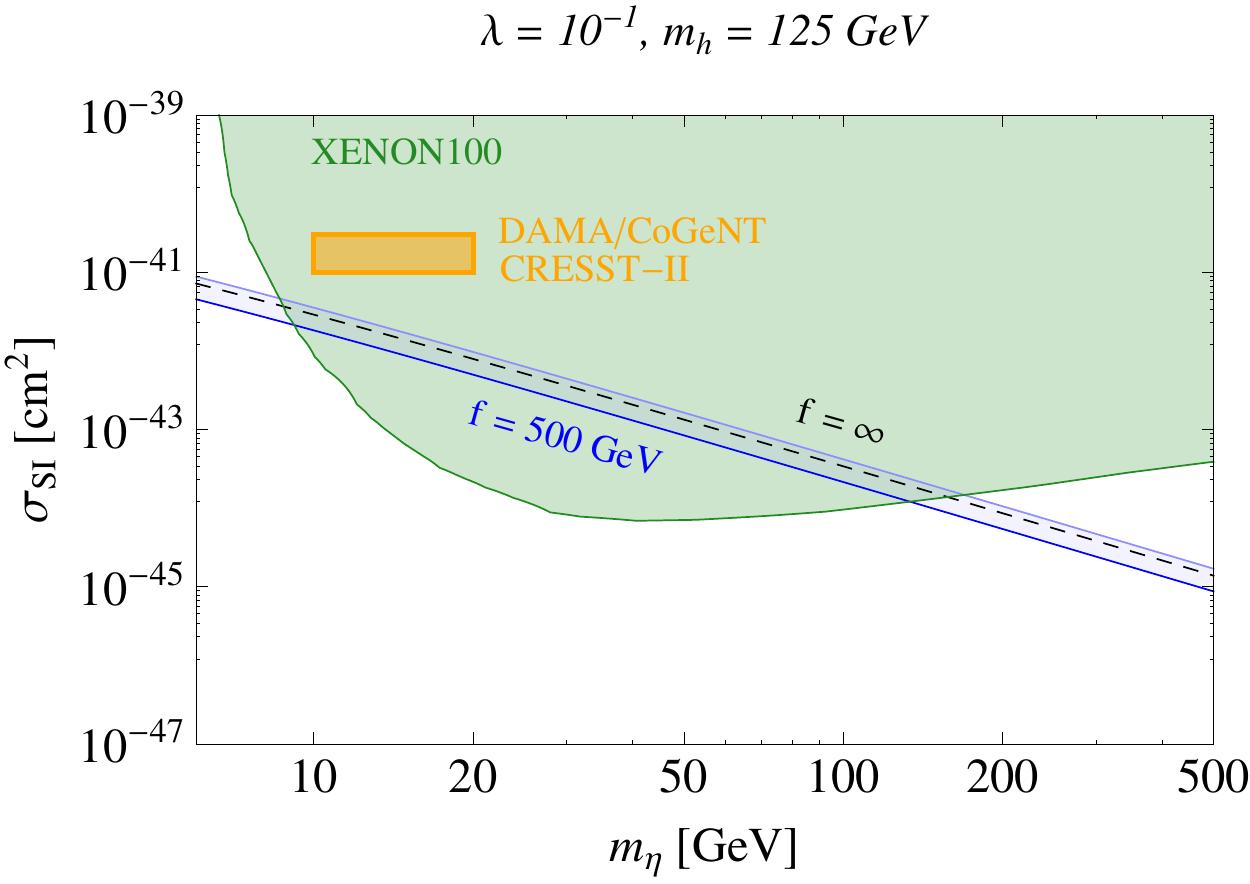}
   \end{minipage}\hspace{1.5 cm}
  \begin{minipage}{0.40\textwidth}
   \centering
   \includegraphics[scale=0.65]{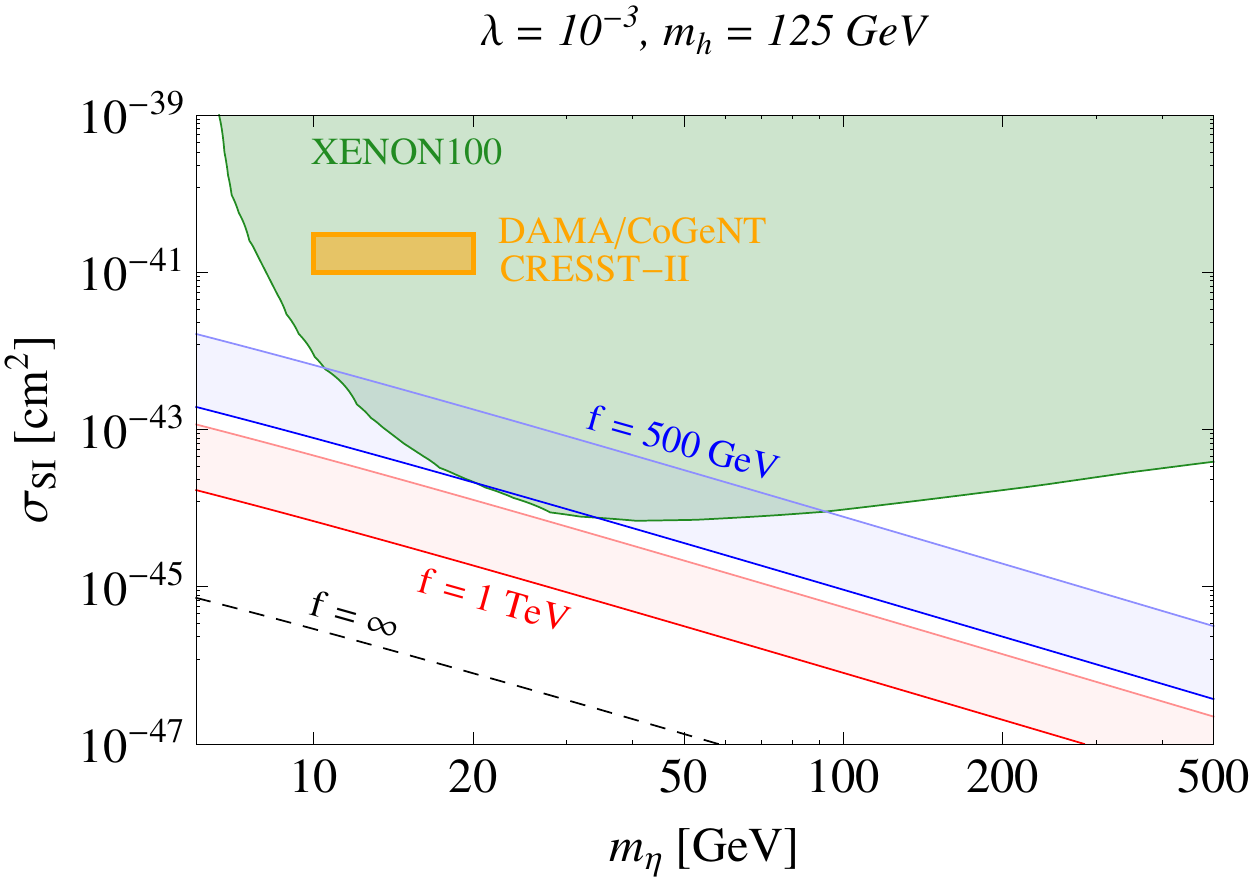}
   \end{minipage}
\caption{\emph{The spin-independent cross section for the elastic scattering of the DM candidate $\eta$ off nuclei. The green shaded region is excluded by XENON100  \cite{Aprile:2011hi}, while the  orange shaded region roughly corresponds to the excess events reported by DAMA, CoGeNT and CRESST-II \cite{Hooper:2012ft}. 
The predictions of our scenario are labeled in the same way as in Fig.~\ref{fig:BoltzmannRelic}. 
The two panels correspond to two different values of $\lambda$, and we took Case 2 for  the DM-quark couplings $c_q$.
In the left panel, the band for $f=1$ TeV (not shown) is very 
similar to the one for $f=500\GeV$. }}
\label{fig:DirectDetectionBound}
\end{figure}

The XENON100  experiment recently set the best experimental upper limit on $\sigma_{\rm SI}$ \cite{Aprile:2011hi}, 
and it plans to considerably improve its sensitivity already within the end of 2012. 
The excluded region at 90 \% C.L. is shaded in green in Fig.~\ref{fig:DirectDetectionBound}.
On the other hand, the DAMA/LIBRA collaboration 
(confirming previous results by DAMA/NaI) has provided the first evidence of  an annual modulation in the event rate, that could be due to DM, 
with $8.9$ standard deviations from the expected background \cite{Bernabei:2010mq}. 
Similar signals have been observed by CoGeNT \cite{Aalseth:2011wp} and by CRESST-II \cite{Angloher:2011uu}, 
which both measure an excess of low-energy nuclear recoil events. CoGeNT also observes a  seasonal variation.
The analysis performed e.g. in Ref.~\cite{Hooper:2012ft} shows that a DM candidate with $m_{DM}\simeq (10- 20)$ GeV and $\sigma_{\rm SI}
\simeq (1-3)\cdot 10^{-41}$ cm$^{2}$ can account for the excess events reported by each of these experiments. This favoured region is shaded in  orange in 
Fig.~\ref{fig:DirectDetectionBound}.  One can appreciate
the  well-known tension between these signals and the exclusion limits of XENON100, that is presently unexplained.

In Fig.~\ref{fig:DirectDetectionBound}
we compare $\sigma_{\rm SI}$ in our scenario with the DM direct detection constraints,
taking the same values of the parameters that we used for the relic density
in Fig.~\ref{fig:BoltzmannRelic}.
The calculation of $\sigma_{\rm SI}$ is a straightforward application of the standard formalism \cite{Jungman:1995df} to our composite scenario, that is defined by the  effective interactions of $\eta$ with quarks and gluons. The detailed equations are reported in appendix \ref{app:B}, and can be summarized by the following estimate, which shows explicitly the relative size of the different contributions:
\begin{align}\label{SIApprox2}
\sigma_{\rm SI} & \simeq 3.5 \cdot 10^{-40} \textrm{cm}^2 \lp\frac{10\GeV}{m_\eta}\rp^2 \lp\frac{125\GeV}{m_h}\rp^4 \nonumber\\
&\times \left[\lambda\left(1-\frac{2v^2}{f^2}\right) + \frac{m_h^2}{f^2} \, {\rm Re}
\left(0.04\, c_u + 0.11\, c_d+ 0.18\, c_s + 0.22 \sum_{q=c,b,t} c_q \right) \right]^2~.
\end{align}
The first term in square brackets comes from  Higgs-exchange, while the second term comes from the direct coupling between the DM and the 
light quarks and gluons (through a loop of heavy quarks) in the nucleon. Contrary to the relic density, the DM direct detection bounds 
are sensitive also to the couplings between $\eta$ and the light quarks, mostly the strange.
The Fig.~\ref{fig:DirectDetectionBound} shows $\sigma_{\rm SI}$ for the Case 2 defined in section \ref{sec:symm}:
we vary  $c_d=c_s=c_b$ in the range defined by \eq{param} and we take $c_{u,c,t}=0$. 
Notice that the effect of the derivative coupling of \eq{eq:L1} is very small at low momentum transfer and it has been neglected.

For $\lambda=10^{-1}$ (left panel of Fig.~\ref{fig:DirectDetectionBound}), 
the first term of \eq{SIApprox2} dominates
and $\sigma_{\rm SI}$ is of the same order of magnitude as in the non-composite case.
By comparing with Fig.~\ref{fig:BoltzmannRelic}, one remarks that the correct relic density can be 
reproduced by $m_\eta \sim 10$ GeV, with a DM candidate that lies close to the region preferred by DAMA, CoGeNT and CRESST-II, or by $m_\eta \sim 80$ GeV,
in a region already excluded by XENON100, or by $m_\eta > 150$ GeV, with no constraints from present direct detection bounds.
For $\lambda=10^{-3}$ (right panel of Fig.~\ref{fig:DirectDetectionBound}), the difference w.r.t. the non-composite case is more important.
The correct relic density is reproduced e.g. for $f=1$ TeV and $m_{\eta}\simeq 100$ GeV (see Fig.~\ref{fig:BoltzmannRelic}). This candidate is compatible with 
the present XENON100 bound, and it is within the reach of near future measurements with improved sensitivity. 
For such small values of $\lambda$, on the contrary, the non-composite case cannot be probed in DM direct detection experiments.

We remark that the various contributions in \eq{SIApprox2} can have opposite sign and (partially) compensate each other, reducing $\sigma_{\rm SI}$. 
Above we assumed conservatively that all contributions are positive, but one cannot exclude a cancellation, e.g. switching to a negative value of $\lambda$.
Therefore the DM direct detection bound is quite sensitive to the details of the theoretical model, and for specific values of the parameters
our scenario can avoid this constraint.

Let us make now a few remarks about DM indirect detection experiments. Bounds from direct detection - as shown in Fig.~\ref{fig:DirectDetectionBound} 
- allow for a light DM candidate with a mass $\lesssim 10$ GeV. In this regime, the DM annihilation in the early Universe is entirely driven by the process 
$\eta\eta\leftrightarrow b\overline{b}$, with a required cross section of the order of $\langle \sigma v_{rel} \rangle\simeq 3\cdot 10^{-26}$ cm$^{3}$s$^{-1}$.
This value weakly depends on the DM relative velocity, and therefore it remains constant from the freeze-out until today, resulting into a sizable antiproton flux on the earth, 
produced by the DM annihilation into quarks that hadronize. We estimated this flux using the tools  in Ref.~\cite{Cirelli:2010xx}. 
As already pointed out in Ref.~\cite{Lavalle:2010yw}, 
this antiproton flux can be larger than the one observed by the PAMELA experiment \cite{Adriani:2010rc}, resulting in a constraint on the DM mass that is 
complementary to the direct detection one. 
One should keep in mind, however, the large astrophysical uncertainty of these observations, mainly due to the propagation of charged particles in the Galaxy. 
A mass $m_\eta$ smaller than 10 GeV is generically  disfavoured, with the exact bound depending on the adopted propagation model.
In addition, the DM annihilations can also leave a trace in the cosmic microwave background power spectra, the size of the effect being proportional to 
$\langle\sigma v_{rel}\rangle/m_\eta$. Assuming that $\langle\sigma v_{rel}\rangle$ is approximately velocity-independent,
one finds that  the Planck satellite should reach a sufficient sensitivity to observe such effect, as long as $m_\eta \lesssim 50$ GeV \cite{Galli:2009zc}.

We remark also that when the DM annihilation process occurs close to the Higgs resonance, 
the Breit-Wigner enhancement mechanism is operative \cite{Feldman:2008xs,Ibe:2008ye}, resulting in a large 
boost of the annihilation cross section today, w.r.t. its freeze-out value. 
This boost factor can be invoked to explain some recent anomalies reported in cosmic ray data \cite{Feldman:2008xs,Guo:2009aj}.  Notice that the same resonant enhancement may play a crucial role
for the pair annihilation of DM particles into two, monochromatic
high-energy gamma rays. This channel is particularly relevant, given
its peculiar spectral structure that cannot be confused with the continuum
astrophysical backgrounds. A value of the DM mass too close to the Higgs
resonance can exceed the upper limit
($\langle\sigma v_{rel}\rangle_{\gamma\gamma}\sim 4\cdot 10^{-28}$cm$^{3}$s$^{-1}$
for $m_{\eta}\simeq 60$ GeV) recently established by the Fermi-LAT
collaboration \cite{Ackermann:2012qk}.

Finally, let us compare the DM direct detection bounds with those coming from collider searches of monojet plus missing transverse energy. Indeed, the singlet $\eta$ might be pair-produced through its coupling to quarks; while  the two singlets escape the detector, these events
can be identified by the initial state radiation of a gluon or a photon.
In some cases \cite{Fox:2011pm} these searches can provide comparable bounds to DM direct detection experiments, in particular for light fermionic DM (with $m_{DM}\lesssim 10$ GeV). This is however not the case for our composite singlet for three reasons. First, the reach of monojet searches is reduced for a light scalar DM  than it is for a fermionic DM \cite{Goodman:2010ku}. Furthermore, in our scenario, the effective coupling of $\eta$ to gluons depends on the momentum 
of the process (at the  momenta involved in DM direct detection experiments,  $\eta$ couples 
more strongly to gluons  than  at collider energies),  contrary  to  Ref.~\cite{Fox:2011pm} where the effective coupling was assumed to be constant. 
Finally, in our case bounds on the invisible Higgs decay (section  \ref{sec:LHC})   provide the strongest  collider constraint.   For these reasons we neglect bounds from monojet searches in what follows.

%%%%%%%%%%%%%%%%%%%%%%%%%%%%%%%%%%%%%%%%%%%%%%%%%%%%%%%%%%%%%%%%%%%%
\section{Combined results}\label{sec:combined}
%%%%%%%%%%%%%%%%%%%%%%%%%%%%%%%%%%%%%%%%%%%%%%%%%%%%%%%%%%%%%%%%%%%%

\begin{figure}[!bt!]
\centering
\includegraphics[scale=0.8]{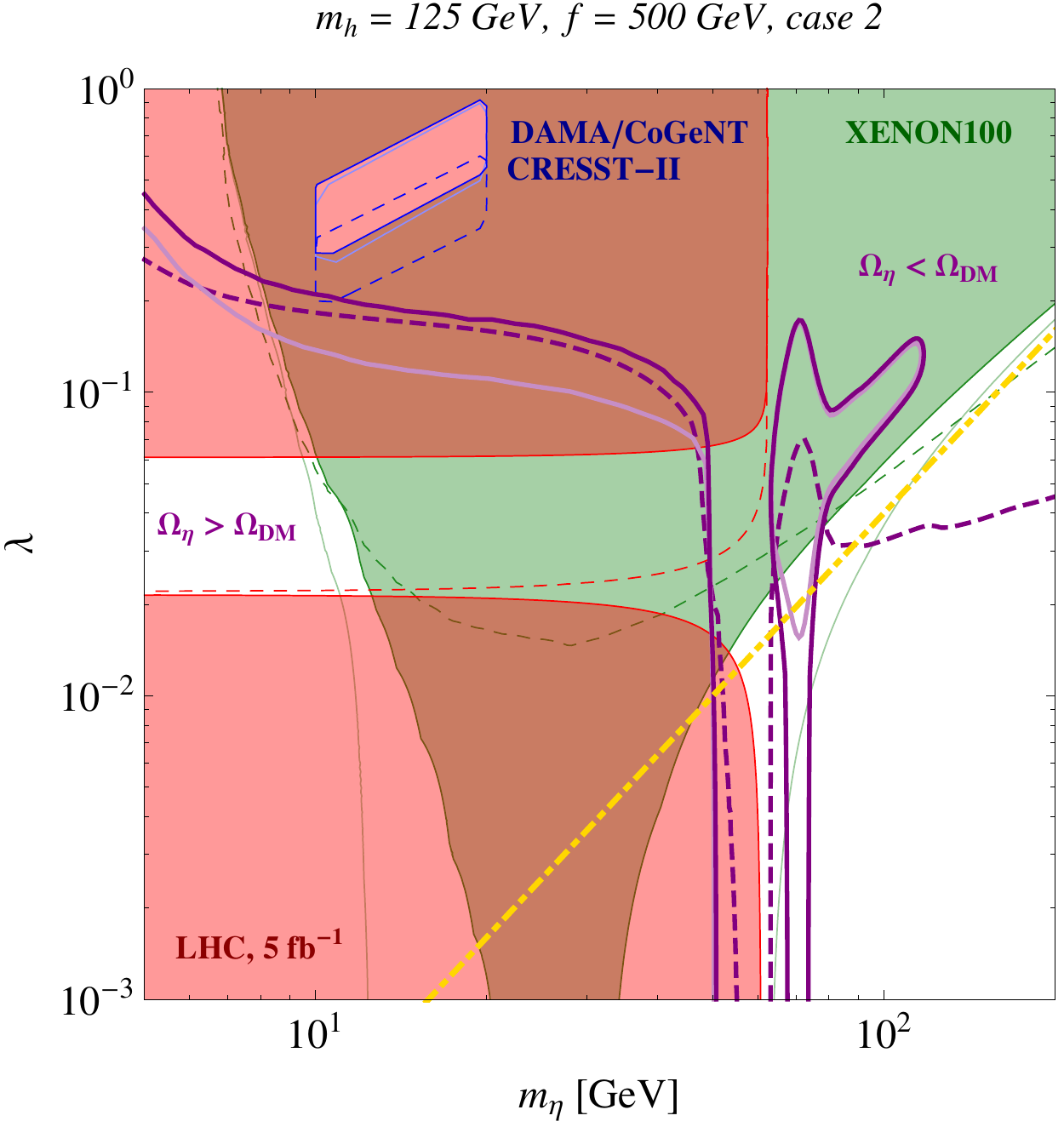}
\caption{\emph{The contour $\Omega_\eta =\Omega_{DM}$ (solid dark purple line)
 in the plane $(m_{\eta},\lambda)$, for $m_h=125$ GeV, $f=500$ GeV,
 assuming  Case 2 with $c_b=1/2$. 
 The green shaded region is disfavoured by XENON100, the region delimited by a blue line 
 is favoured by DAMA/CoGeNT/CRESST-II,
 and the red shaded region is disfavoured by the Higgs signal at the LHC.
 The solid light purple/green/blue lines correspond to the same observables for maximal $c_b$ ($a=b=1$ in \eq{param}).
 The dashed purple/green/blue/red lines correspond to the same observables in the non-composite case, $f=\infty$.
 Finally, the region below the yellow dot-dashed line corresponds to the  theoretical preferred region defined by \eq{expectTheo}.}}
\label{fig:ExclusionPlot2}
\end{figure}

\begin{figure}[!htb!]
\begin{minipage}{0.4\textwidth}
   \centering
   \includegraphics[scale=0.66]{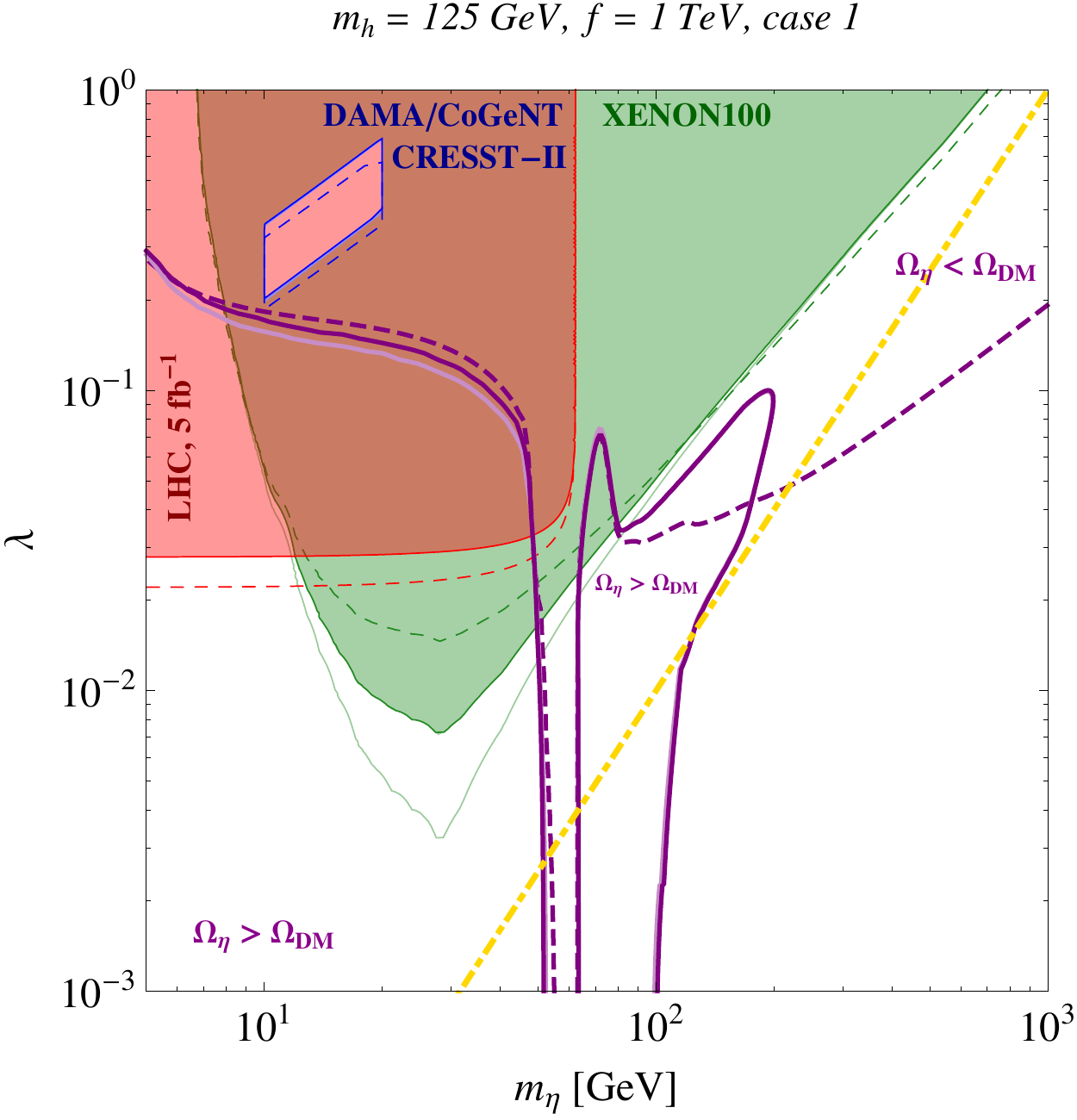}
   \end{minipage}
    \hspace{1.5 cm}
   \begin{minipage}{0.4\textwidth}
    \centering
    \includegraphics[scale=0.66]{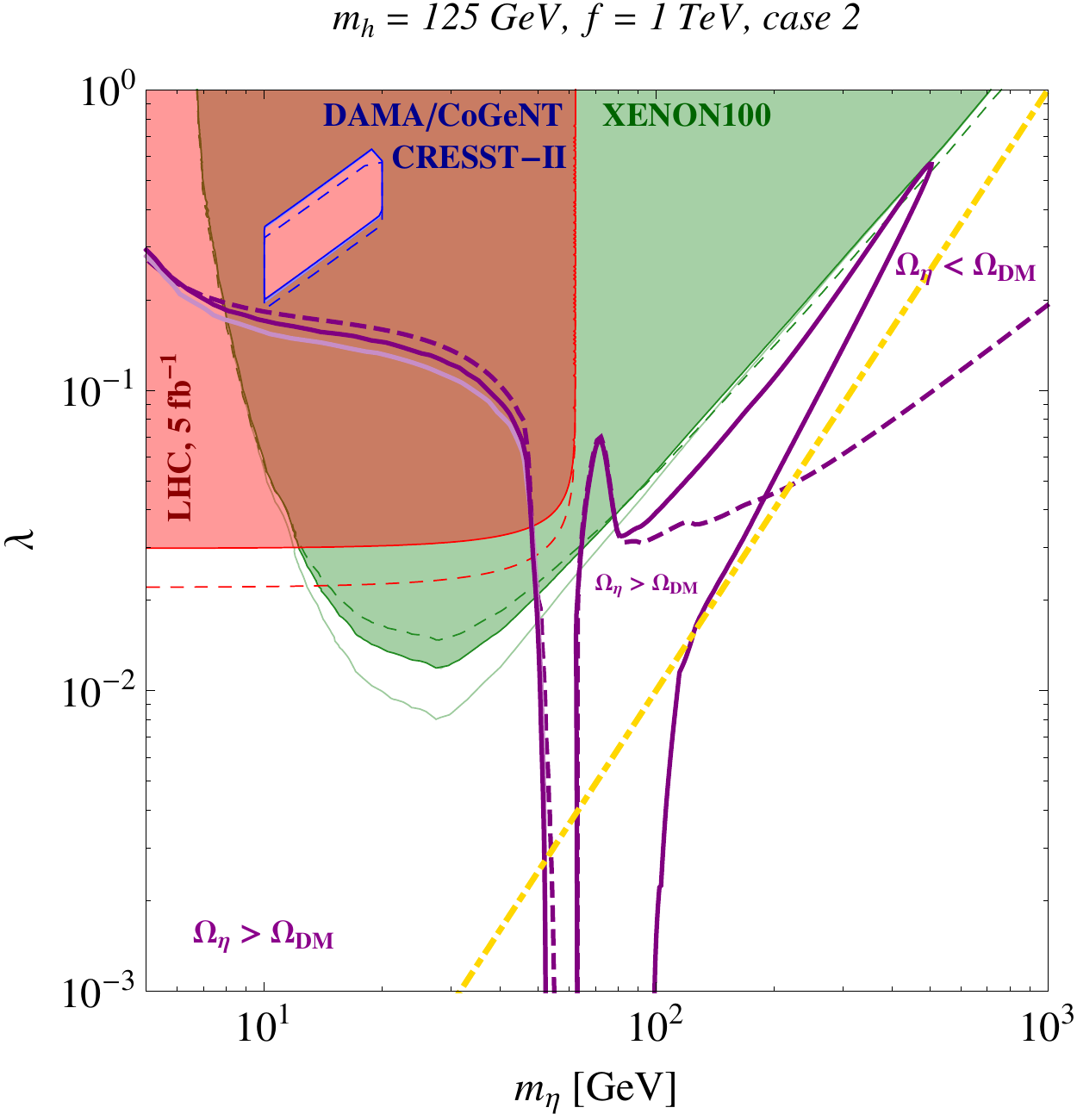}
    \end{minipage}
 \caption{\emph{The same as in Fig.~\ref{fig:ExclusionPlot2}, but with $f=1$ TeV and with a comparison between two scenarios for the top quark couplings: Case 1 (left panel) 
 and Case 2 (right panel), as defined at the end of section \ref{sec:symm}.}} 
 \label{fig:ExclusionPlot1}
\end{figure}

\begin{figure}[!tb!]
   \centering
   \includegraphics[scale=0.8]{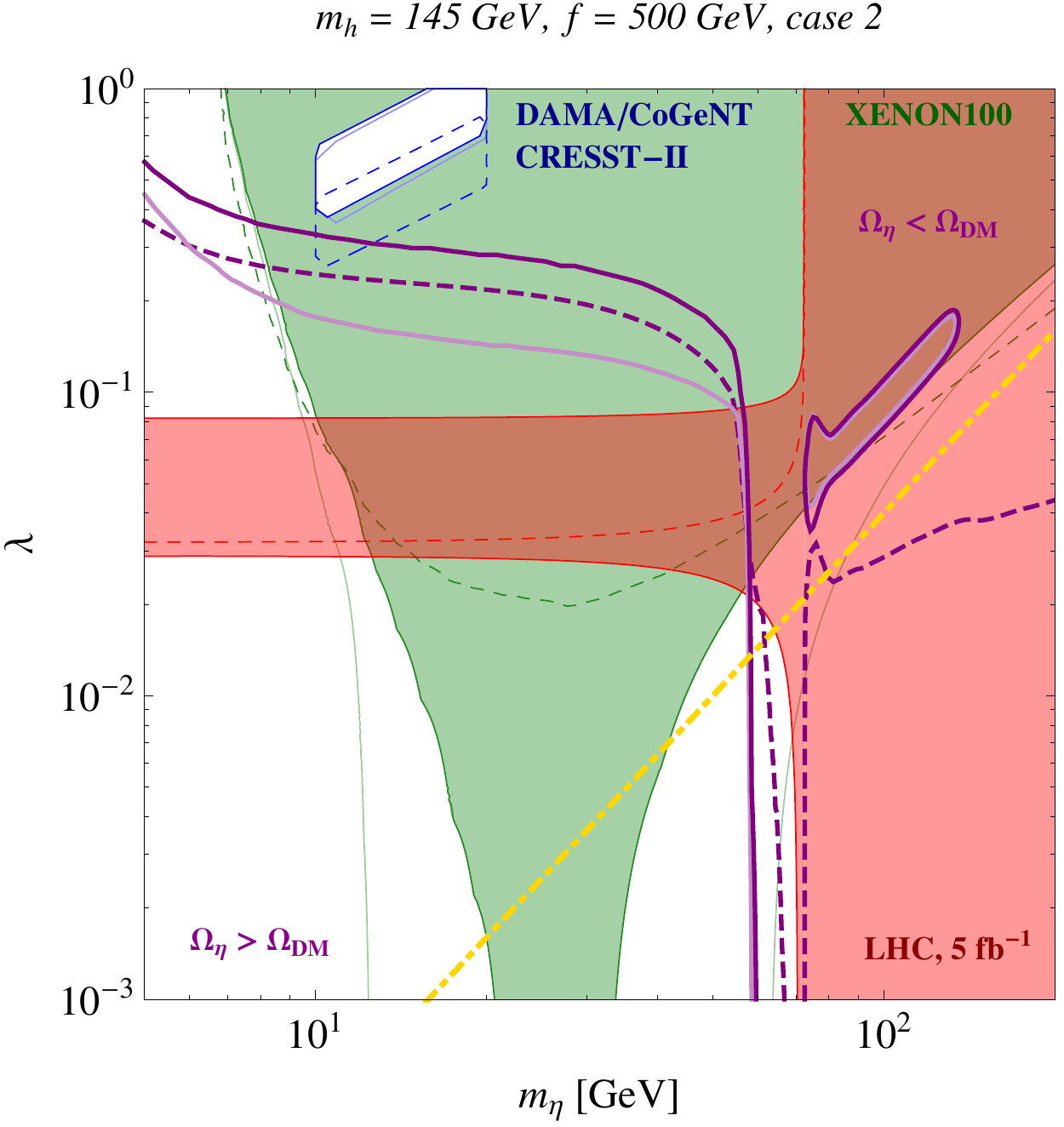}
 \caption{\emph{
The same as in Fig.~\ref{fig:ExclusionPlot2}, but with $m_h=145$ GeV.}}
 \label{fig:ExclusionPlot3}
\end{figure}

We present in this section a combined analysis of the parameter space for the composite singlet DM candidate $\eta$.
The results are displayed in Figs.~\ref{fig:ExclusionPlot2}-\ref{fig:ExclusionPlot3}, in the plane $(m_\eta , \lambda)$. 
We require to reproduce the observed value of the DM relic density (the purple contour corresponds to $\Omega_\eta=\Omega_{DM}$), 
and we take into account DM direct detection experiments (the green region is disfavoured by XENON100)
and Higgs searches (the red region is disfavoured by ATLAS and CMS).\\

\noindent \underline{$\bullet$ $m_h=125$ GeV, $f=500$ GeV: a DM candidate with $m_{\eta}\simeq 70$ GeV}

We begin by taking the Higgs mass value presently preferred at the LHC, $m_h=125$ GeV, and choosing the characteristic pNGB scale
close to the lower bound coming from electroweak precision tests, $f=500$ GeV. We focus on Case 2 
(a vanishing coupling $c_t$
between the top quark and the DM), since it  allows for a light $\eta$, $m_\eta \lesssim m_h$. This scenario is illustrated in 
Fig.~\ref{fig:ExclusionPlot2}.

In the non-composite case ($f= \infty$, dashed contours), 
a viable DM candidate lies in correspondence to the Higgs resonance ($m_{\eta}\simeq 60$ GeV, $\lambda\lesssim 0.02$), 
or beyond the kinematical threshold for annihilation into electroweak gauge bosons ($m_{\eta} \gtrsim 80$ GeV, $\lambda\simeq 0.04$).

In the composite case the situation changes considerably.
Note first that the region $m_{\eta}\lesssim m_h/2$ is disfavoured by the LHC Higgs signal excess, because
the Higgs decay width is dominated by the hidden decay channel into DM,
except for  $\lambda\simeq m_h^2/(2f^2)\simeq 0.03$, where a cancellation in $\Gamma_{inv}$ takes place.
Note also that the direct detection bound is strengthened by the composite interactions, because of the contribution of
direct DM$-$quark couplings, see \eq{SIApprox2}.

Taking into account these constraints, a definite prediction follows from Fig.~\ref{fig:ExclusionPlot2}: $m_\eta\simeq 70$ GeV
with $\lambda \lesssim 0.02$. 
At small values of $\lambda$, the $\eta$-$\eta$-$h$ vertex
is dominated by the derivative coupling, therefore the relic density becomes independent from $\lambda$ and the 
purple lines  in  Fig.~\ref{fig:ExclusionPlot2} become vertical.
For higher DM masses the derivative interaction becomes too strong to accommodate the relic density
(unless one enters in the ``cancellation region", at relatively large values of $\lambda$, which is however excluded by XENON100).

One may ask if the region of parameters that is allowed phenomenologically is also compatible with the theoretical expectations.  Independently of the specific model, we expect from  \eq{expectTheo} that $m_\eta^2 \gtrsim \lambda f^2$. 
The region satisfying this relation lies below the yellow dot-dashed line in  Fig.~\ref{fig:ExclusionPlot2},
and it is compatible with the  phenomenologically preferred region.
\\

\noindent\underline{$\bullet$ $m_h=125$ GeV, $f=1$ TeV: DM candidates with $m_{\eta}\simeq 60$ GeV and $100\lesssim m_{\eta}\lesssim 500$ GeV}

As the scale $f$ increases, the composite interactions become weaker, and the bounds from the LHC Higgs signal and from XENON100 become less stringent
and closer to the non-composite case.
This is illustrated in Fig.~\ref{fig:ExclusionPlot1}, where we take $f=1$ TeV. 
In particular, all values of $m_\eta$ are viable for $\lambda\lesssim 10^{-2}$.

The correct DM relic density
can be accommodated  for   $m_\eta$ lying  a bit below or above the Higgs resonance
 at  $m_h/2\sim  60$ GeV, or for $m_{\eta}\gtrsim 100$ GeV,  where
the derivative  interaction $\eta$-$\eta$-$h$ in \eq{eq:L1} becomes of the right order 
to give the  correct annihilation cross-section above the $WW$ threshold. 
Furthermore, for relatively large values of $\lambda$, one enters in the cancellation region described in section \ref{Sec:RelicDensity}: the DM annihilation is suppressed and  the
relic density can be accommodated even for very large values of the DM mass, up to $m_\eta\simeq 500$ GeV in Case 2 (right panel of Fig.~\ref{fig:ExclusionPlot1}).
If the annihilation into $t\bar{t}$ is stronger (Case 1, left panel of Fig.~\ref{fig:ExclusionPlot1}), the allowed region closes earlier, at $m_\eta \simeq 200$ GeV.

As discussed above, composite models prefer $\lambda \lesssim m_\eta^2/f^2$
(the region below the yellow dot-dashed line) 
that  is compatible with the Higgs-resonance region for $\lambda \lesssim 0.003$,
and with the region dominated by the derivative coupling, for $\lambda \lesssim 0.02$. 
On the contrary, the cancellation region is slightly disfavoured theoretically, even though
$\lambda$ larger by a factor of a few is sufficient to realize the cancellation.
\\

\noindent\underline{$\bullet$ $m_h=145$ GeV, $f=500$ GeV: DM candidate with  $m_{\eta}\lesssim 10- 20 \GeV$ and $m_\eta\simeq 60$ GeV}

In case the LHC excess at 125 GeV were not confirmed, the Higgs boson might be heavier, as long as it decays invisibly with a sufficient rate to avoid
the LHC bound. 
In order to illustrate this possibility, in Fig.~\ref{fig:ExclusionPlot3} we choose a representative value $m_h=145$ GeV, 
assuming for definiteness $f=500$ GeV and Case 2
(the results are very similar in Case 1). 

The LHC bound is satisfied easily below the kinematical threshold for Higgs decays into DM,  $m_{\eta}<m_{h}/2$,
in a region of parameters which is very much complementary to the one allowed in Fig.~\ref{fig:ExclusionPlot2}.
As a consequence, a light DM candidate with a mass $m_{\eta}\lesssim10$ GeV is compatible with XENON100 and LHC.
A light singlet with $10 \GeV \lesssim m_\eta \lesssim 20 \GeV$ could in principle
explain the DAMA/CoGeNT/CRESST-II results (of course, in tension with
the XENON100 bound). However the value $\lambda\lesssim 0.3$ required by the relic density is slightly smaller than the one needed to fit the
signal in these experiments: for larger $\lambda$, $\eta$ accounts only for part of the DM relic density.

In any case, the large coupling $\lambda \gtrsim 0.1$, needed to explain the relic density when $m_\eta\lesssim 20$ GeV, 
is in contradiction with theoretical expectations.
Indeed, the singlet receives at least a contribution to its mass-squared of order
$\lambda v^2\gtrsim (80 \GeV)^2$,
and the NDA expectation is actually much larger, $m_\eta^2\gtrsim \lambda f^2\gtrsim (160\GeV)^2$. 
Therefore, the case of a light singlet is disfavoured in the context of composite models, since it requires a large cancellation 
between different contributions to the singlet mass.

The other viable DM candidate in Fig.~\ref{fig:ExclusionPlot3} lies just before the Higgs resonance, at $m_\eta\simeq 60$ GeV, as long as $\lambda\lesssim 0.02$. 
This solution lies in the theoretically favoured region, and it requires a not too large $\textrm{Re}(c_b)$, in order to avoid the XENON100 bound.
\\

In summary, it is  interesting that the composite scalar DM model is very predictive, despite  the new  non-renormalizable interactions, 
generated by the strong sector at the TeV scale. In some cases, the composite singlet DM is even more constrained  than the 
non-composite one. 
Indeed, in good approximation, the relic density and the LHC bounds depend only on 
$m_\eta$, $\lambda$ and the derivative interaction between $\eta$ and the
Higgs, whose strength is uniquely fixed by the pNGB scale $f$.
Therefore, our results are independent from the specific composite model, up to the relatively small modifications due to the choice of the parameters $c_t$ 
(compare the two panels in Fig.~\ref{fig:ExclusionPlot1}) and $c_b$ (compare the dark and light contours in Figs.~\ref{fig:ExclusionPlot2}-\ref{fig:ExclusionPlot3}).
The largest model-dependence lies in the DM direct detection bound for small values of $\lambda$.
The most interesting DM candidates (for $m_h=125\GeV$) 
have mass $m_\eta\simeq 60- 70 \GeV$, around the Higgs resonance, or lie in the region $100 \GeV \lesssim m_\eta \lesssim 200 \GeV$. 
In both cases small values of the Higgs portal coupling, $\lambda\lesssim 0.03$, are favored both phenomenologically and theoretically.

%%%%%%%%%%%%%%%%%%%%%%%%%%%%%%%%%%%%%%%%%%%%%%%%%%%%%%%%%%%%%%%%%%%%%%%%%%
\section{Conclusions and prospects}\label{sec:conc}
%%%%%%%%%%%%%%%%%%%%%%%%%%%%%%%%%%%%%%%%%%%%%%%%%%%%%%%%%%%%%%%%%%%%%%%%%%

We have studied the phenomenology of a  composite DM  candidate, arising from  the $O(6)/O(5)$ global symmetry breaking pattern. 
The light scalar spectrum of this model is formed by five pNGBs, which correspond to the Higgs doublet $H$ and the  real singlet $\eta$, 
that plays the role of the DM. 
They are lighter than the dynamically-generated  scale $m_\rho\sim $ TeV,
due to their pNGB nature.
We have shown  that the DM phenomenology is strongly affected  by the composite nature of the pNGBs,
that allows for
\begin{enumerate}[(i)] 

\item  derivative interactions between
$\eta$ and $H$ which grow as $p^2/f^2$, where $p$ is the relevant momentum;

\item  contact interactions $c_\psi$ between $\eta$ and the SM fermions $\psi$ that break
the $\eta$-shift symmetry  and are suppressed by $m_\psi p/f^2$;

\item    modification of the Higgs couplings to gauge bosons and fermions  at order $v^2/f^2$.

\end{enumerate}
We found that this scenario is very predictive, and it is actually easier to test than the non-composite case, which depends only
on the Higgs portal interaction $\lambda$.

The most important phenomenological consequences of  the properties {\rm (i)-(iii)} are  the following.
The $\eta$ annihilation cross section grows with $m_\eta$ because of the derivative interaction between $H$ and $\eta$. Therefore, to reproduce the
DM relic density, one needs $m_\eta\lesssim 500$ GeV (for $f\lesssim 1$ TeV). 
The same derivative interaction implies that the invisible Higgs decay in two DM particles is possible even for vanishing $\lambda$. 
On the other hand, the composite Higgs can be slightly fermiophobic and privilege decays into gauge bosons, as preferred by recent LHC data.
The couplings $c_\psi$ can significantly increase the DM scattering cross-section on nuclei: a signal in DM direct detection 
experiments is possible even for vanishing $\lambda$, as long as $10$ GeV $\lesssim m_\eta\lesssim 100$ GeV and $f\lesssim 1$ TeV.

Once all the constraints are taken into account, we are left with only three regions of parameters for our DM candidate:
\begin{enumerate}

\item  Light mass region, with $m_\eta \simeq 10- 20 \GeV$, close to the values preferred by the signals of DAMA/CoGeNT/CRESST-II.
In this case the Higgs has a substantial invisible width into DM, that would be strongly disfavoured
if the LHC hint for $m_h\simeq 125\GeV$ were confirmed. Such light DM also contrasts with the theoretical expectations, because it requires a large coupling
$\lambda\gtrsim 0.1$, and therefore a strong cancellation is needed to make the DM mass small.

\item Intermediate mass region, with $m_\eta\simeq 50- 70 \GeV$, where $\eta$  annihilates mostly through the Higgs resonance.
This region of parameters is almost  the same as in the non-composite case, but
a discrimination is possible for $f\sim 500$ GeV, where the LHC and DM direct detection constraints are significantly different. For $m_\eta<m_h/2$, Higgs invisible decays are expected, resulting in a reduction of the LHC Higgs signal.

\item Heavy mass region, with $m_\eta\simeq 100- 500 \GeV$, in which 
the annihilation cross-section is dominated by the derivative interaction between $H$ and $\eta$. The latter is fixed by the value of $f$, 
and the relic density is actually too small for $f\sim500$ GeV, but it becomes the correct one when $f\sim 1\TeV$. 
This case is possibly the most interesting, since the relevant DM interactions are completely determined by its composite nature, and 
it can be tested in near future DM direct searches (in contrast to the non-composite case).
Moreover a DM mass close to the Higgs mass would strongly suggest a common origin, that is to say,
both are generated radiatively by order one couplings to the top quark.

\end{enumerate}

Other  indirect  signals of our scenario are the following. 
Since the Higgs is also composite, its couplings to the SM particle
differ from an ordinary Higgs, as it can be  explicitly seen  in Eqs.~(\ref{hVV})-(\ref{hpp}).
  It has been shown in Refs.~\cite{Giudice:2007fh,Espinosa:2010vn}
  that  the LHC with greater integrated luminosity 
  could discriminate between standard and composite Higgs 
for values of $f\sim 500 $ GeV.
Furthermore, the  Higgs can decay  into two $\eta$ that  could be measured as
an  invisible  decay width, as long as $\eta$ is sufficiently lighter than the Higgs. 
When $\eta$ is heavier, at large center-of-mass energy searches for monojets plus missing transverse energy 
might be the most sensitive to the derivative coupling, and they deserve further investigation.
Also, heavy resonances $\rho$  of the strong sector  
could decay  into   the DM  with a sizable branching ratio for large $g_\rho$.
For example, the decay $\rho\rightarrow \eta \eta W$  
could result in a  distinctive signal  with leptons and large missing energy in the final state.

Let us briefly comment on another cosmological implication of a singlet scalar. It has been pointed out \cite{Espinosa:2011eu} that a real singlet with the parity of  \eq{z2new} 
can help in inducing a strongly first order electroweak phase transition, and thus play an important role in electroweak baryogenesis, as well as leave 
an observable spectrum of gravitational waves \cite{Kehayias:2009tn}. It would be interesting to check whether this possibility can be realized together with the singlet being DM. 
Although 
it seems unrealistic for an elementary singlet (the couplings needed to fit the relic density are too small to play a role during the phase transition),
it could be possible for a composite $\eta$, that has large additional interactions.

In summary, we have presented a minimal composite framework that accounts for the Higgs boson and for a scalar DM in a natural way.
Future DM searches (e.g. XENON1T) and LHC measurements shall be able to close on the three most promising regions of parameters 
described above,
and thus establish or refute the connection between the EWSB sector and the DM sector.

%%%%%%%%%%%%%%%%%%%%%%%%%%%%%%%%%%%%%%%%%%%%%%%%%%%%%%%%%%%%%%%%%%%%%%%%%%
%%%%%%%%%%%%%%%%%%%%%%%%%%%%%%%%%%%%%%%%%%%%%%%%%%%%%%%%%%%%%%%%%%%%%%%%%%

\begin{center}
{\bf Acknowledgments}
\end{center}
We thank Roberto Contino, Jose Miguel No, Ennio Salvioni, Javi Serra, and Geraldine Servant for useful discussions. 
This work was  partly supported by the project FPA2011-25948. 
The work of AU was also supported  by the Emergence-UPMC-2011 research program,
MF was also supported  by the  Marie-Curie Reintegration Grant PERG06-GA-2009-256374 within the European Community FP7,
and AP was also  supported by the project 2009SGR894 and ICREA Academia Program.
AU and MF  thank the IFAE (Barcelona) for hospitality during the first part of this project.

%%%%%%%%%%%%%%%%%%%%%%%%%%%%%%%%%%%%%

%%%%%%%%%%%%%%%%%%%%%%%%%%%%%%%%%%%%%%%%%%%%%%%%%%%%%
\appendix

\section{The composite $O(6)/O(5)$ model for dark matter}\label{sec:theory}
%%%%%%%%%%%%%%%%%%%%%%%%%%%%%%%%%%%%%%%%%%%%%%%%%%%%%

In this appendix we present a model of a composite sector that accommodates the Higgs doublet $H$ and the DM singlet $\eta$
as pNGBs. 
The analysis is intended to provide the reader with a concrete realization of the effective lagrangian
that we used all over our phenomenological study,  and to compute the expected values of the pNGBs masses and couplings. 
The coset $SO(6)/SO(5)$ provides
the minimal realization of a set of pNGBs with the quantum numbers of $H$ and $\eta$.
This coset was first studied in Ref.~\cite{Gripaios:2009pe}, with  focus on the modifications of the Higgs 
phenomenology in the presence of a real scalar singlet.
Here we demand that $\eta$ plays the role of DM particle and we derive its properties, under the requirement
of a consistent  EWSB.

We assume  a strong sector, with a mass gap $m_\rho\sim $ TeV, whose
global symmetry breaking pattern is
$SO(6) \rightarrow SO(5)$. This  can be achieved by a composite field $\Sigma$ in the fundamental representation \textbf{6}
of $SO(6)$, that acquires a VEV
\begin{equation}\label{SigmaVEV}
\Sigma_0=(0,0,0,0,0,1)^T~.
\end{equation}
The five  NGBs transform  as  a ${\bf 5}$ of $SO(5)$, which decomposes as
$\mathbf{4} \oplus \mathbf{1} \simeq (\mathbf{2},\mathbf{2}) \oplus (\mathbf{1},\mathbf{1})$
under  the subgroup $SO(4) \simeq SU(2)_L \times SU(2)_R$.
The NGBs describe the fluctuations along the broken directions, whose generators can be written as
\begin{equation}
T^{\hat a}_{ij}  = -\frac{i}{\sqrt{2}}
 \left( \delta^{\hat a}_i \delta^6_j - \delta^{\hat a}_j \delta^6_i \right),~~~~{\hat a}=1,\dots,5.
\label{brogen} 
\end{equation}
The $SO(6)/SO(5)$ coset is parametrized by
\begin{align}
\label{Sigma}
\Sigma=\exp\left(i\frac{\sqrt{2}\pi^{\hat a}T^{\hat a}}{f}\right) \Sigma_0
&= \sin\frac{\pi}{f} 
\left( \frac{\tilde{h}_1}{\pi},~\frac{\tilde{h}_2}{\pi},~\frac{\tilde{h}_3}{\pi},~\frac{\tilde{h}_4}{\pi},~\frac{\tilde{\eta}}{\pi},~ 
\cot\frac{\pi}{f} \right)^T \nonumber\\
&=\left(h_1,~h_2,~h_3,~h_4,~\eta,~\sqrt{1-h^2-\eta^2}\right)
~,
\end{align}
with $\pi^{\hat a} \equiv (\tilde{h}_i,\tilde\eta)$, $\pi\equiv\sqrt{\sum_{i=1}^4\tilde{h}_i^2+\tilde\eta^2}$,
and a field redefinition
\begin{equation}
h_i \equiv \frac{\tilde{h}_i}{\pi}  \sin\frac{\pi}{f} ~,~~~~~ 
\eta \equiv \frac{\tilde{\eta}}{\pi}  \sin\frac{\pi}{f} ~,
\label{def}
\end{equation}
with $h^2\equiv\sum_{i=1}^4 h_i^2$.
The usual Higgs doublet is given by $H=f[(h_1+ih_2)/\sqrt{2},\ (h_3+ih_4)/\sqrt{2}]$.
In the unitary gauge three NGBs are eaten by the weak gauge bosons, and one is left with 
the Higgs boson $h\equiv h_3$ and $\eta$.

The NGB chiral lagrangian reads, up to terms with four derivatives,
\bea
\nonumber
{\cal L}_{kin} = \frac{f^2}{2} |D_{\mu} \Sigma|^2 &=& \frac{f^2}{2} \left[ (\partial_\mu h)^2
+(\partial_\mu\eta)^2 
+\frac{(h\partial_\mu h+\eta\partial_\mu\eta)^2}{1-h^2-\eta^2} \right. \\ \label{kin}
&+& \left.\frac{g^2h^2}{2}\left(W^{\mu+} W^-_\mu+\frac{1}{2\cos^2\theta_W}Z^\mu Z_\mu\right)\right]\, .
\label{lkina}
\eea
From the last term in \eq{kin}, it is clear that the scale of EWSB is set by the VEV of $h$: $\langle h \rangle \equiv v/f=\sqrt{\xi}$.
Note that ${\cal L}_{kin}$ is symmetric under the parity
$P_\eta:~\eta\rightarrow -\eta$,
but higher derivative terms in the $SO(6)/SO(5)$ chiral lagrangian, involving the Levi-Civita tensor
(such as the Wess-Zumino-Witten term \cite{Gripaios:2009pe}), are not.
Since $P_\eta$ is crucial to make $\eta$ stable and thus a viable DM candidate,
we need to assume that it is a symmetry of the whole composite sector and it is not spontaneously broken.
This amounts to take the global symmetry breaking pattern to be $O(6)\rightarrow O(5)$,
with 
\be
P_\eta={\rm diag}(1,1,1,1,-1,1)\, ,\label{peta}
\ee
 and to require $\langle \eta \rangle =0$.
Note also that, once $h$ acquires a VEV, its kinetic term receives a correction from the third term in the square bracket of \eq{lkina}.
The physical Higgs and DM bosons (with a canonical kinetic term) are defined as follows:
\be
\frac{h_{phys}}{f} = \frac{h-\langle h \rangle}{\sqrt{1-\xi}}~,\qquad \frac{\eta_{phys}}{f} =\eta~.
\label{phys}\ee
In all the observables we studied, we consistently used the couplings of the physical fields
(but dropping the subscript `$phys$' everywhere).

The EWSB is triggered by the couplings of the SM gauge bosons and fermions to the composite sector, that 
break $O(6)$ explicitly, generating an effective potential for the pNGBs at the one-loop level.
Still, we will show that under some conditions $P_\eta$ is preserved by the gauge and fermion interactions as well.
In this case, the most general form of the effective potential for $h$ and $\eta$
can be written as
\begin{equation}
V_{eff}(h,\eta) =\frac{f^2}{2} \left(\mu_h^2 h^2 + \mu_\eta^2\eta^2 \right)+  \frac{f^4}{4}\left(\lambda_h h^4 + 2 \lambda h^2 \eta^2 +\lambda_\eta\eta^4\right)
 + O((h,\eta)^6)~.
\label{Veff}\end{equation}
Such potential must be minimized taking into account the constraint 
$h^2+\eta^2 = \sin(\langle\pi\rangle/f) \le 1$, 
that follows from \eq{def}.
We look for a minimum with $0< h=v/f < 1$, to realize EWSB, and with $\eta = 0$, to preserve DM stability. 
The necessary and sufficient
conditions for this to be a local minimum are $0<-\mu_h^2<f^2 \lambda_h$ and $\mu_\eta^2 \lambda_h > \mu_h^2 \lambda$, where we neglected
the dimension-six terms in \eq{Veff}.\footnote{
The requirement for this to be the global minimum (in the region $h^2+\eta^2 \le 1$) involves some extra
lengthy conditions, that are satisfied in a wide range of the potential parameters.} 
Then, the pNGB physical masses are given by
\be
m_h^2 \simeq -2 \mu_h^2 = 2 \lambda_h v^2~,~~~~~ m_\eta^2 \simeq  \mu_\eta^2 + \lambda v^2 ~,
\label{mhe}\ee
up to $O(\xi)$ corrections.

The size of the effective potential coefficients depends on the way the SM gauge bosons and fermions couple to the composite sector.
Note that, in order to reproduce the hypercharge of the SM fermions, the global symmetry of the composite sector should be enlarged to
$O(6)\times U(1)_X$, with the hypercharge defined by 
\begin{equation}\label{Hypercharge}
Y\equiv T^3_R+X ~.
\end{equation}
Since $\eta$ is a gauge singlet, its NGB nature is not affected by the gauging of $SU(2)_L\times U(1)_Y$, that is to say, it does not
acquire a potential through the  gauge loops.  Therefore, the electroweak gauge bosons only contribute to the Higgs potential,
inducing $\mu_h^2 \sim g^2m_\rho^2/(16\pi^2)$ and $\lambda_h \sim g^4/(16\pi^2)$ (see Ref.~\cite{Gripaios:2009pe} for details). 
These contributions are generally smaller than those coming from the top quark, which will drive EWSB.

The Eqs.~(\ref{lkina}), (\ref{phys}) and (\ref{Veff}) define all the relevant interactions
among $\eta$, $h$ and the gauge bosons. The corresponding Feynman rules are displayed in Table \ref{fig:FRulesCDM}.

Coming to the SM fermion interactions with the strong sector,
we assume the partial compositeness scenario \cite{PartComp,CH:ACP},
which is preferred by the constraints on flavour violation:
each SM chiral fermion $\psi$ couples linearly to a composite operator ${\cal O_\psi}$
of the strong sector:
\be
{\cal L}_{int}=\lambda_\psi  \overline{\psi}\, {\cal O_\psi}+h.c.\, . 
\label{intfer}
\ee
This leads to a mixing  of the SM fermions  with the heavy composite states (of mass  $\sim m_\rho$)
 of order $\lambda_\psi/g_\rho$. 
By repeating this reasoning for each SM chiral fermion, the Yukawa couplings turn out to be given by 
$y_\psi\simeq \lambda_{\psi_L}\lambda_{\psi_R}/g_\rho$.

Following standard techniques \cite{CH:ACP,Mrazek:2011iu},  we will  promote the SM fields $\psi$ into  spurions $\Psi$
 forming   complete $SO(6)$ representations.
The $\Psi$  are  defined to transform as  the   $SO(6)$-multiplet  ${\cal O_\psi}$  
in such a way that  the interaction \eq{intfer}  can be written as an  invariant under $SO(6)$.
The interactions between the SM fermions and the pNGBs $h$ and $\eta$ will crucially depend  on the
 representation of  $\Psi$  under $SO(6)$. 
We will identify the simplest representations that (i) allow to generate the Yukawa coupling, (ii) induce a pNGB effective potential that realizes EWSB satisfactorily,  (iii) preserve the parity $P_\eta$ that guarantees the stability of our DM candidate $\eta$.

In addition to these necessary requirements, the choice of the $SO(6)$ representation for $\Psi$ also determines
whether $\lambda_\psi$ breaks or preserves the $\eta$ shift symmetry $U(1)_\eta$, that is to say, whether or not $\psi$-loops contribute to the effective potential for $\eta$. This is crucial to determine the DM mass.
To study this issue, it is useful to decompose the $SO(6)$ multiplets under the maximal 
subgroup $SO(4)\times SO(2)_\eta \simeq SU(2)_L\times SU(2)_R \times U(1)_\eta$, where
$SO(2)_\eta\simeq U(1)_\eta$ is precisely the symmetry associated with the NGB $\eta$, that is generated by $T_\eta=T^{\hat 5}$ defined in \eq{brogen}:
\begin{equation}
T_\eta=
\frac{1}{\sqrt 2}\left(
\begin{array}{cc}
 0_{4\times4} & 0_{4\times2}     \\
 0_{2\times4}  & \sigma_2  \\
\end{array}
\right)\, .
\label{teta}
\end{equation}
From \eq{peta} and \eq{teta} we have
\be
[P_\eta,T_\eta]\not=0\, ,
\ee
therefore one cannot assign to the  SM fields   a definite $P_\eta$ parity and a 
non-zero $U(1)_\eta$ charge at the same time.
This means that both symmetries can be preserved by the SM couplings 
to the strong sector only if the SM fields  transform trivially under the $SO(2)_\eta$, i.e. they
are not charged under $U(1)_\eta$.

Let us classify the spurions $\Psi$ that can accommodate the SM fermions while preserving $P_\eta$.
The SM fermion isosinglets (isodoublets) can be embedded in any $SO(6)$ representation that contains a singlet (doublet) of $SU(2)_L$.
We find that an embedding preserving $P_\eta$ is possible using a vector representation, $\Psi \sim {\bf 1},~{\bf 6},~{\bf 15},~ {\bf 20}^\prime$, \dots,
or a pair of spinor representations, ${\bf 4} + \overline{\bf 4}$, ${\bf 10} + \overline{\bf 10}$, and so on  (when acting on spinors, $P_\eta$ interchanges conjugate representations).
Their  $SU(2)_L\times SU(2)_R\times U(1)_\eta$ decomposition reads
\bea
  \textbf{4}&=  &   (\textbf{2},\textbf{1})_{+1}\oplus (\textbf{1},\textbf{2})_{-1} ~,\nonumber  \\
  \textbf{6}&=  &  (\textbf{2},\textbf{2})_0\oplus(\textbf{1},\textbf{1})_{+2}\oplus(\textbf{1},\textbf{1})_{-2} ~,\nonumber  \\
  \textbf{10}&=   &  (\textbf{2},\textbf{2})_{0}\oplus (\textbf{3},\textbf{1})_{+2}\oplus (\textbf{1},\textbf{3})_{-2} ~, \nonumber\\
  \textbf{15}&=  &  (\textbf{1},\textbf{3})_{0}\oplus(\textbf{3},\textbf{1})_{0}\oplus(\textbf{1},\textbf{1})_0 \oplus(\textbf{2},\textbf{2})_{+2} \oplus(\textbf{2},\textbf{2})_{-2} ~,\nonumber \\
  \textbf{20}^\prime&=   &  (\textbf{3},\textbf{3})_0\oplus(\textbf{2},\textbf{2})_{+2}\oplus(\textbf{2},\textbf{2})_{-2}\oplus(\textbf{1},\textbf{1})_{+4}\oplus(\textbf{1},\textbf{1})_{-4}
  \oplus(\textbf{1},\textbf{1})_0  ~.
  \label{61520}
\eea
The components that can contain the SM fermions and do not transform under 
$U(1)_\eta$ are, for the isodoublets, 
the $(\textbf{2},\textbf{2})_0$ in the $\textbf{6}$ or in the $\textbf{10}$,
and for the isosinglets either  the $(\textbf{1},\textbf{1})_0$ or  the 
$(\textbf{1},\textbf{3})_0$, which are present  in the $\mathbf{1}$, the $\mathbf{15}$ or the $\mathbf{20}^\prime$. 
These results are summarized in Table \ref{tab:reps}.

\begin{table}[t]
\begin{center}
\begin{tabular}{|c||c|c|c|c|c|c|c|c|}
\hline
$\psi$  $\backslash$ $\Psi$ &\textbf{1}&\textbf{4}+$\overline{\textbf{4}}$
&\textbf{6}&\textbf{10}+$\overline{\textbf{10}}$
&\textbf{15}&\textbf{20$^\prime$}& \dots\\
 \hline
 \hline
$\begin{array}{c} q_L,l_L\\ u_R,d_R,e_R \end{array}$  & $\begin{array}{c}  -\\ \surd  \end{array}$& $\begin{array}{c}\times \\ \times  \end{array}$
&$\begin{array}{c} \surd \\\times  \end{array}$&$\begin{array}{c} \surd \\\times  \end{array}$
&$\begin{array}{c}\times \\ \surd  \end{array}$
&$\begin{array}{c}\times \\ \surd  \end{array}$
&$\begin{array}{c}\dots \\ \dots \end{array}$\\
\hline
\end{tabular}
\end{center}
\caption{\emph{$SO(6)$ representations  for the spurions $\Psi$ that can embed the SM fermions $\psi$ preserving $P_\eta$. 
The symbol $\surd$ ($\times$)
indicates that the $P_\eta$-preserving embedding  does not break (does  break) the $SO(2)_\eta$ subgroup.}\label{tab:reps}}
\end{table}

In the following we will describe the properties of two models.
In the first one we will embed all the SM fields in the $\bf 6$ of $SO(6)$, and therefore
the right-handed couplings to the strong sector will break the $U(1)_\eta$ symmetry explicitly.
The  top quark loops will give the largest contributions to all terms of the pNGB effective potential in  \eq{Veff}; this corresponds to Case 1 of section \ref{sec:symm}.
In the second model, on the other hand, we will embed the up-type quark singlets, and in particular $t_R$, in the $\bf 15$,
so that  $U(1)_\eta$ will be  preserved by the top couplings; doublets and down-type quark singlets will be embedded in the $\bf 6$ as before.
In this case the DM particle can be lighter than the EW scale leading to  a substantially different phenomenology; this corresponds to Case 2. 
Both models satisfy the properties (i)-(iii) given above.

Before going to these two specific models, let us briefly mention an  alternative way  to keep
$\eta$ light.
If the   SM embedding  in $SO(6)$ representations preserves the
$SO(5)$ subgroup, that is broken by the Higgs VEV   down to an $SO(4)'$, we would  
have after EWSB four exact  NGBs, the  three  SM  ones eaten by the weak gauge bosons plus $\eta$.
In this case the mass of $\eta$  would be  protected by this $SO(4)'$ symmetry
and not by the  $SO(2)_\eta$ of \eq{teta}. 
It is clear, however,  that the SM fermion doublets cannot be embedded in
 complete $SO(5)$ representations, so $\eta$ gets always a mass  from loops of this sector.
A relative light $\eta$ could be achieved  by requiring  the smallest possible coupling
of the fermion doublets to the strong sector. Here we do not explore this possibility any further.

\begin{table}
\begin{center}
  \includegraphics[width=16.5 cm]{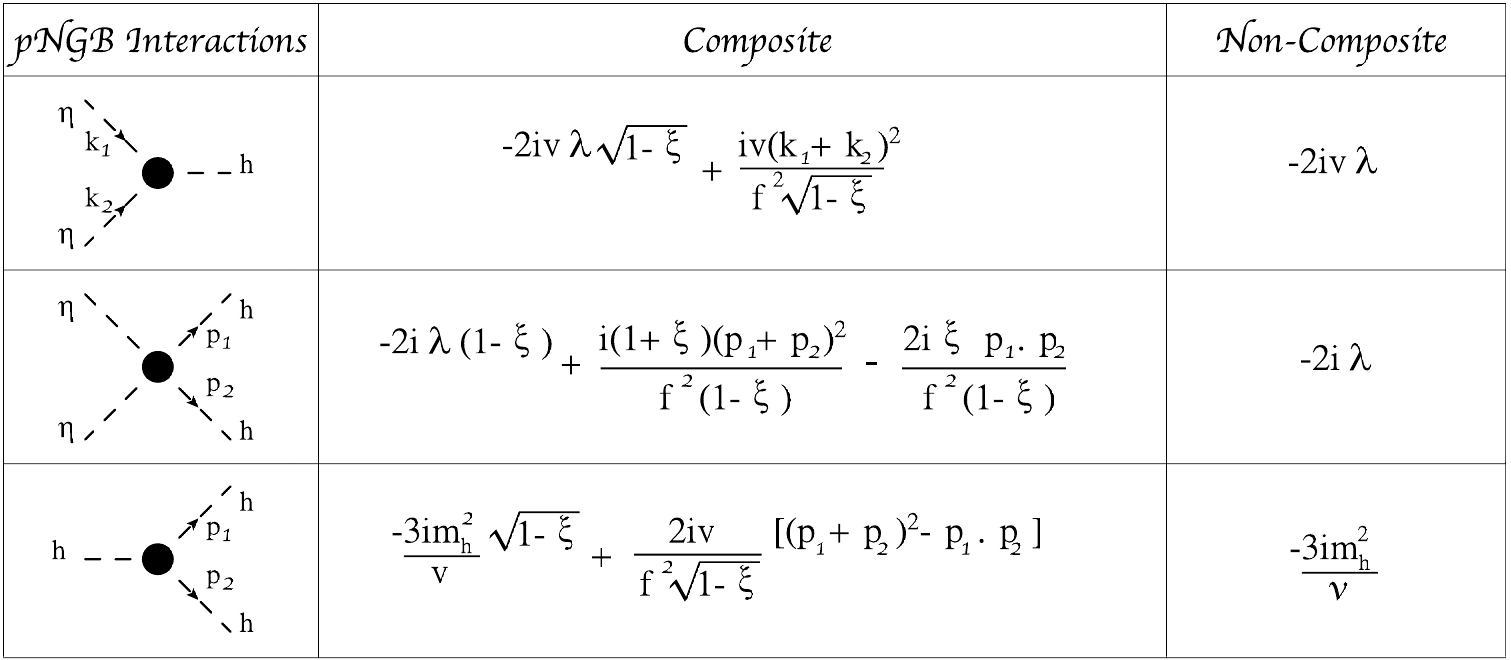}
    \includegraphics[width=16.5 cm]{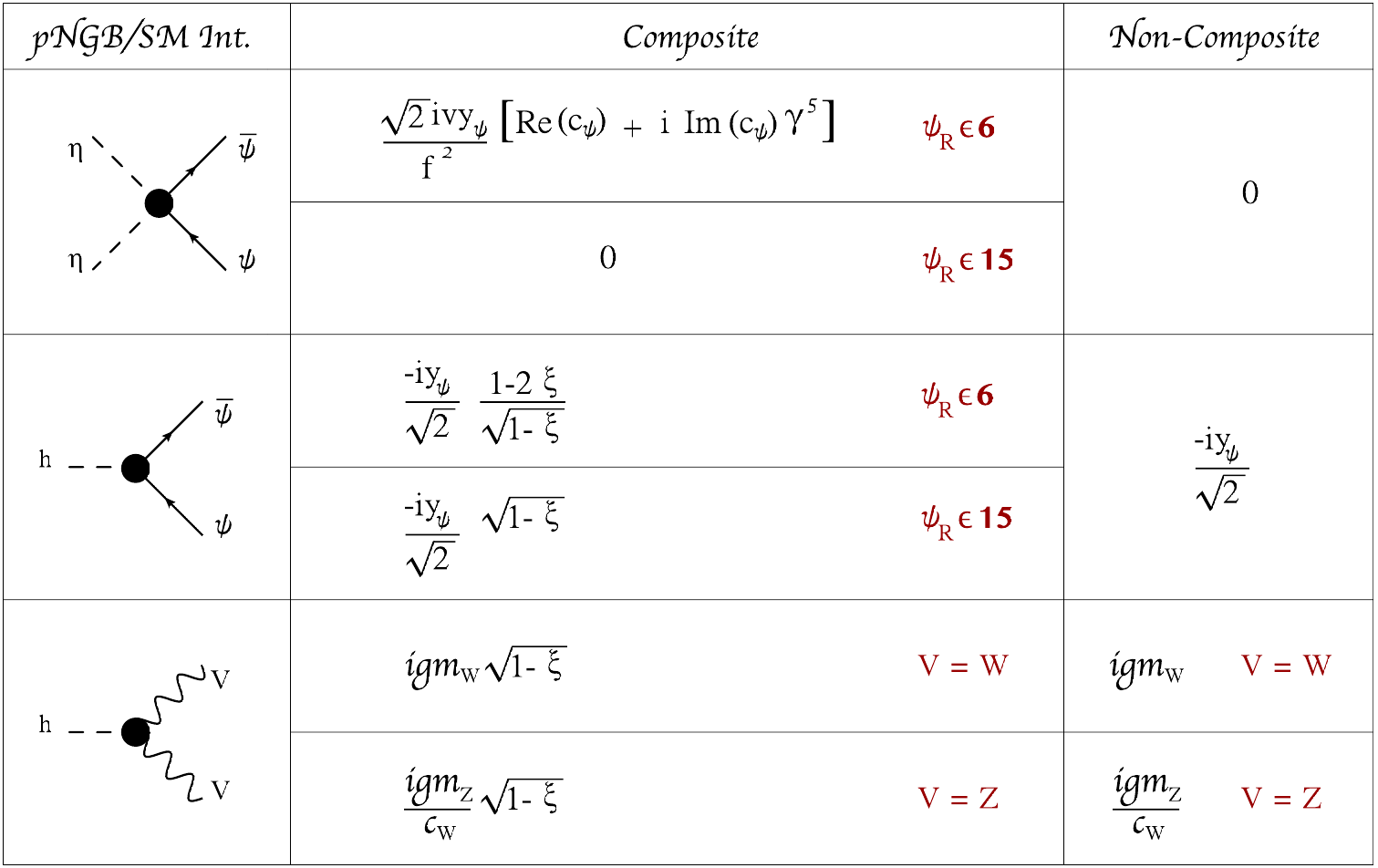}
  \caption{\emph{Feynman rules for the relevant interactions between the Higgs boson $h$ and the DM particle $\eta$ (upper panel) and for the interactions of $h$ and $\eta$ with the SM fermions and gauge bosons (lower panel). The second column corresponds to the composite model with coset $O(6)/O(5)$, while the third column corresponds to the SM plus a real scalar singlet.}}\label{fig:FRulesCDM}
  \end{center}
\end{table}

\subsection{Case 1: all the SM  fermions in the    representation ${\bf 6}$}
%%%%%%%%%%%%%%%%%%%%%%%%%%%%%%%%%%%%%%%%%

Let us begin with embedding all SM fermions into   $\Psi \sim {\bf 6}$ \cite{Gripaios:2009pe}.
A SM electroweak doublet can be embedded in the the bi-doublet with $P_\eta=+1$, 
while for a SM isosinglet the only  embedding with $P_\eta=+1$ (to avoid the breaking of $P_\eta$) 
is given by
$\Psi_{\psi_R}=(0,0,0,0,0,\psi_R)$.
Coming to the $U(1)_X$ assignments, recall that the pNGB field $\Sigma$ does not carry $X$-charge. 
Using \eq{Hypercharge}, the isosinglet up-type  and down-type  quarks   should respectively be embedded  into a spurion with  $X_{u_R}=2/3$  and $X_{d_R}=-1/3$. 
Then, in order to allow for the Yukawa couplings to $\Sigma$, the isodoublet quarks  $q_L$  should be embedded into  two  spurions, 
one with $X_{q_L}=2/3$ to generate  the up-quark Yukawa, $y_u\simeq \lambda_{q_L}\lambda_{u_R}/g_\rho$, and
one with $X_{q_L^\prime}=-1/3$ to generate  the down-quark Yukawa, $y_d\simeq\lambda_{q_L'} \lambda_{d_R}/g_\rho$.\footnote{ 
In the case of the third family, the coupling $\lambda_{q_L'}$ implies a correction to the SM value of the coupling  $Zb_L\overline{b_L}$, 
because it corresponds to  $T_R^3(b_L)=1/2$ different from $T_L^3(b_L)=-1/2$ \cite{Agashe:2006at};
we assume a small value $\lambda_{q'_L} \sim y_b$,
to guarantee a small enough correction to $Zb_L\overline{b_L}$.}
Here and in the following we tacitly assume that the lepton couplings to the composite sector are generated in analogy with the quark ones.

As can be seen from the $U(1)_\eta$ charges of the ${\bf 6}$ components in \eq{61520}, the $\eta$ shift symmetry is broken by the embedding of the right-handed quarks.
Therefore $\eta$ will receive an effective potential from loops involving only  isosinglet quarks.
This can be seen explicitly by writing all the invariants that involve the quarks and the pNGBs.
The techniques of Refs.~\cite{CCWZ,Mrazek:2011iu}
are useful to count the number of possible invariants,
 that we write using  the spurion notation
\begin{align}
\label{fermionhiggs}
& (\bar\Psi_{q_L} \Sigma) (\Sigma^T \Psi_{u_R}) =  \frac{1}{\sqrt{2}}\bar u_L u_R\, h \sqrt{1-\eta^2-h^2} \, ,\nonumber\\
& (\bar\Psi_{q_L^\prime}\Sigma) (\Sigma^T \Psi_{d_R}) =-\frac{1}{\sqrt{2}} \bar d_L d_R\, h \sqrt{1-\eta^2-h^2}\, ,\nonumber \\
& (\bar{\Psi}_{u_R} \Sigma) \pslash (\Sigma^T\Psi_{u_R} ) =  \bar u_R \pslash  u_R (1-\eta^2-h^2) \, ,\nonumber\\
& (\bar{\Psi}_{d_R}  \Sigma ) \pslash (\Sigma^T\Psi_{d_R} ) =  \bar d_R \pslash  d_R (1-\eta^2-h^2) \, ,\nonumber\\
&  ( \bar{\Psi}_{q_L} \Sigma ) \pslash (\Sigma^T\Psi_{q_L}  ) =  \frac{1}{2}\bar u_L  \pslash u_L\, h^2\, , \nonumber\\ 
&(\bar{\Psi}_{q_L'}  \Sigma) \pslash (\Sigma^T\Psi_{q_L'}) =\frac{1}{2}  \bar d_L  \pslash d_L\, h^2\, .
\end{align}
We can now justify the parameterization of  \eq{param} for  the DM$-$bottom coupling. 
The first and second lines in \eq{fermionhiggs}  generate the quark masses and, at order $\eta^2$, they also 
give a contribution to the couplings of \eq{eq:L2},
$\Delta c_u=\Delta c_d=1/2$, up to ${\cal O}(\xi)$ corrections.
The third and fourth lines in \eq{fermionhiggs} correspond to derivative couplings analog to the one in \eq{psidirectcoupling},
that can be cast in the form of \eq{eq:L2} by using the Dirac equation.
Using NDA to estimate the coefficients of these invariants, one obtains
$\Delta c_{u,d} = O(\lambda_{u_R,d_R}^2/g_\rho^2) + i\, O(\lambda_{u_R,d_R}^2/g_\rho^2)$. Note that the ratio $\lambda_{u_R,d_R}^2/g_\rho^2$
varies approximately in the range $[y_{u,d}^2/g_\rho^2,1]$, depending on the relative size of $\lambda_{q_L,q_L'}$ and $\lambda_{u_R,d_R}$.
Adding the two contributions to the DM$-$fermion couplings, one recovers \eq{param}.\footnote{
In general, the contributions of order $\lambda_{u_R,d_R}^2/g_\rho^2$ could have negative sign and partially compensate
the contribution $1/2$. In our analysis we did not consider the possibility of  such a cancellation.}

The Feynman rules for the interactions of $\eta$ and $h$ with fermions,
following from \eq{fermionhiggs}, are reported in Table~\ref{fig:FRulesCDM}.

The effective lagrangian in momentum space, obtained after integrating out the composite sector, can be written as
\begin{align}
\label{fermionsL}
\mathcal{L}_{\rm f} =& \sum_{r=q_L,u_R,q_L^\prime,d_R} \Big[\Pi_0^r(p^2) (\bar{\Psi}_r \pslash\, \Psi_r) 
- \Pi_1^r(p^2) (\bar{\Psi}_r \Sigma)  \pslash  (\Sigma^T \Psi_r )\Big]  \nonumber \\
- & \left[ M^u(p^2) (\bar \Psi_{q_L} \Sigma ) (\Sigma^T\Psi_{u_R}  ) +M^d(p^2) (\bar\Psi_{q_L^\prime}\Sigma  ) (\Sigma^T\Psi_{d_R} )+ h.c.\right] \, ,
\end{align}
where $p$ is the momentum of the fermion fields. 
Parametrically, at small momentum ($p \lesssim m_\rho$) we have
\be
\Pi_0^r\simeq 1, \qquad \Pi_1^{r}\simeq \frac{\lambda_{r}^2}{g_\rho^2}~,\qquad
M^u \simeq\frac{\lambda_{q_L}\lambda_{u_R}}{g_\rho} f~,\qquad
M^d\simeq \frac{\lambda_{q_L'}\lambda_{d_R}}{g_\rho} f~.
\label{fermionsBC}
\ee
Eqs.~({\ref{fermionhiggs}) and (\ref{fermionsL}) imply that the fermion masses read
$m_{u,d} = M^{u,d}(0) /\sqrt{2} (v/f)\sqrt{1-\xi}$.
Loops of SM fermions contribute to the Coleman-Weinberg potential for the pNGBs. Since the heaviest fermions give the dominant contribution, 
we concentrate on the third family of quarks. At one-loop, from \eq{fermionsL} one obtains 
\begin{align}\label{Pottot}
V^{({\bf 6})}_{t,b}(h,\eta) = - 2N_c \int \frac{d^4 p}{(2\pi)^4} &\left\{ 
 \log\left[ p^2 \lp  \Pi_0^{q_L}+\Pi_0^{q_L^\prime}  - \frac{\Pi_1^{q_L}}{2}  h^2\rp \left( \Pi_0^{t_R} - \Pi_1^{t_R}  (1-\eta^2-h^2)\right) \right.\right.\nonumber\\
&\left.\left. - \frac{|M^t|^2}{2}  h^2 (1-\eta^2-h^2) \right] + \log[(q_L,t_R) \leftrightarrow (q'_L,b_R)]
\right\}~,
\end{align}
where the overall negative sign comes from the fermion loop, the factor of 2 from the fermion polarizations and $N_c$ is the number of colors.

For a momentum larger than the mass of the resonances $m_\rho$, one expects the form factors $\Pi_1^{r}$ and $M^t$ to fall off  fast enough to make the integral in \eq{Pottot}  convergent.   
At $p\lesssim m_\rho$, for $\lambda_r\ll g_\rho$ 
one can expand the logarithm in  \eq{Pottot},
since  each power of $h^2$ and $\eta^2$ is associated with a suppression factor  $(\lambda_r\lambda_{r'}/g_\rho^2)$.
One can easily check that all terms in the expansions in $h^2$ and $\eta^2$ receive contributions 
from the top couplings $\lambda_{q_L}$ and $\lambda_{t_R}$, therefore one can safely neglect the terms in $\lambda_{q_L'}$ and $\lambda_{b_R}$, that we assume to be  of the order of the bottom Yukawa. 
The first logarithm in \eq{Pottot} provides the dominant contribution, coming from the top-quark, to the effective potential $V_{eff}$ 
defined in \eq{Veff}. By expanding this contribution and integrating over momenta, we obtain
\bea
\mu_h^2 & = & \frac{ N_c m_\rho^2}{16 \pi^2} \left(c_1 \lambda_{q_L}^2 - 2 c_2 \lambda_{t_R}^2 - 2c_3 y_t^2 \right)
\nonumber ~, \\ 
\lambda_h & = & \frac{N_c}{16\pi^2} \left( 4 c_3 y_t^2 g_\rho^2 +\frac 12  c_1^{(2)} \lambda_{q_L}^4 +  2 c_2^{(2)} \lambda_{t_R}^4 \right)
~, \nonumber \\ 
\mu_\eta^2 & = & \frac{N_c m_\rho^2}{16 \pi^2} \left(- 2 c_2 \lambda_{t_R}^2\right) 
~, \nonumber \\
\lambda_\eta &=& \frac{N_c}{16\pi^2} \left(2c_2^{(2)} \lambda_{t_R}^4 \right)
~,\nonumber \\
\lambda &= & \frac{N_c}{16 \pi^2} \left(2c_3 y_t^2 g_\rho^2+2c_2^{(2)}\lambda_{t_R}^4 \right) ~,
\label{V6exp}
\eea
where $c_i$ and $c_i^{(2)}$ are order one coefficients that account for the uncertainty in the integrals over the form factors.
More specifically, the $c_i^{(2)}$ come from the second order in the log expansion, and therefore they involve an integral over a higher power
of the form factors, 
that in some explicit models is suppressed (see e.g.~Ref.~\cite{hep-ph/0612048}); in the following we
neglect for simplicity these terms, when compared to the $c_i$ ones.

In order for $\xi = -\mu_h^2 / (\lambda_h f^2)$  to be smaller than one, a partial cancellation is needed between the terms in $\mu_h^2$,
that requires $\lambda_{t_R} \sim \lambda_{q_L}$.\footnote{
We are  neglecting   gauge contributions that are smaller than the top one.}
Then, from \eq{V6exp}, using $y_t \simeq \lambda_{q_L}\lambda_{t_R}/g_\rho$ and \eq{mhe},
the Higgs mass can be written  as \begin{equation}
m_h^2= c_3
\frac{ N_c}{2\pi^2}\ y_t^2 \ \xi \ m_\rho^2 \simeq  c_3 (250\ {\rm GeV})^2 \ \frac{\xi}{0.1}  \left(\frac{m_\rho}{2\ \rm TeV}\right)^2
~.
\end{equation}
The coupling $\lambda$ is generated by the same term that controls also $\lambda_h$, so we expect it to be positive and of the same order
\begin{equation}
\lambda\simeq  \frac{\lambda_h}{2}=  \frac{m_h^2}{4v^2}\, 
\end{equation}
(for $m_h=125\GeV$ this corresponds to $\lambda \simeq 1/16$). 
The singlet mass can be written as
\begin{equation}
m_\eta^2 \simeq \mu_\eta^2 = c_2\frac{ N_c}{8\pi^2}\ \lambda_{t_R}^2 \  m_\rho^2\simeq   m_h^2 \frac{c_2}{c_3}   \frac{\lambda_{t_R}^2}{4 y_t^2 \xi}
~.
\label{meta6}\end{equation}
Since the last factor in \eq{meta6} is larger than one, $\eta$ is generically heavier than the Higgs.

%%%%%%%%%%%%%%%%%%%%%%%%%%%%%%%%%%%%%%%%%%%%%%%%%%
\subsection{Case 2: the right-handed top quark in the representation ${\bf 15}$}
%%%%%%%%%%%%%%%%%%%%%%%%%%%%%%%%%%%%%%%%%%%%%%%%%%

As mentioned above, if $t_R$  is embedded into an  $SO(6)$ representation  preserving $U(1)_\eta$, the DM $\eta$ is naturally lighter than the Higgs.
In this section we provide an example of such a model, where up-type isosinglet  quarks are embedded in the $\bf{15}$.\footnote{
 Embedding $t_R$ in the $\bf 1$ of $SO(6)$ is the simplest possibility for light DM, but it is not satisfactory for EWSB,
because in this case $t_R$ does not break the Higgs shift symmetry either. 
As a consequence, one needs to invoke a cancellation in the strong sector to obtain $v\ll f$, and moreover the Higgs mass tends to be too small.}
As  in the previous section we keep all isodoublets and down-type isosinglets in the $\bf{6}$. 
The up-type Yukawa can be generated when  $u_R$  is embedded 
in the ${\bf (1,3)_0}\subset {\bf 15}$. Taking for definiteness $X_{q_L}=2/3$, in the up-type quark sector \eq{fermionhiggs}  is replaced by
\begin{align}
\label{fermionhiggs15}
& \bar\Psi_{q_L} \Psi_{u_R} \Sigma =  \frac{1}{2\sqrt{2}} \bar u_L u_R\, h\,  ,
\nonumber\\
&   \Sigma^T \bar\Psi_{u_R} \pslash  \Psi_{u_R} \Sigma  =  \frac{1}{4} \bar u_R \pslash u_R\, h^2\,  ,
\nonumber\\
&  (\bar{\Psi}_{q_L} \Sigma) \pslash (\Sigma^T \Psi_{q_L} ) =   \frac{1}{2}\bar u_L  \pslash u_L\, h^2\, .
\end{align}
As expected, no terms involving $\eta$ are generated, therefore 
the DM$-$top coupling $c_t$ of  \eq{eq:L2} vanishes,  and there are no
loop contributions to the effective potential for $\eta$ from the up-quark sector.

There are still, however, DM couplings to the down-type quarks;
the heaviest is  the bottom, therefore \eq{meta6} is replaced by
\be
m_\eta^2 \simeq \mu_\eta^2 = c'_{2} \frac{N_c}{8\pi^2} \lambda_{b_R}^2 m_\rho^2\, ,
\ee
which results in a DM mass  $m_\eta \gtrsim O(10)\GeV$ ($\lambda_{b_R} \gtrsim y_b \simeq 1/40$). 
Notice that, by embedding also the $b_R$ in the $\bf{15}$, 
we could make the  DM  even lighter.
Note also that the DM$-$Higgs coupling $\lambda$ is correlated to the DM mass, 
\begin{equation}
\frac{m_\eta^2}{\lambda f^2}\simeq  \frac{g_\rho^2}{{\rm max}(\lambda_{b_L}^2,\lambda_{b_R}^2)} \gg 1~,
\end{equation}
in agreement with the general expectation of \eq{expectTheo}.
Therefore, the lightest the DM is, the weakest its Higgs portal interaction becomes.

Similarly to the previous model,  starting from \eq{fermionhiggs15}
one can compute the top quark contribution to the effective potential,
\begin{equation}
V^{({\bf 15})}_{t}(h) = - 2N_c \int \frac{d^4 p}{(2\pi)^4}  
 \log\left[ p^2 \lp  \Pi_0^q  - \frac{\Pi_1^q}{2}  h^2\rp \left( \Pi_0^t - \frac{\Pi_1^t}{4} h^2 \right)
 - \frac{|M^t|^2}{8}  h^2 \right]  
~. \label{Pot15}
\end{equation}
Expanding and integrating the logarithm, one finds
\bea
\mu_h^2 & = & \frac{ N_c m_\rho^2}{16 \pi^2} \left(c_1 \lambda_{q_L}^2 +\frac 12 c_2\lambda_{t_R}^2 - 2c_3 y_t^2 \right)
\nonumber ~, \\ 
\lambda_h &= & \frac{N_c}{16\pi^2} \left(\frac 12  c_1^{(2)} \lambda_{q_L}^4 + \frac 18 c_2^{(2)} \lambda_{t_R}^4+ c_3^{(2)} y_t^4 \log\frac{m_\rho^2}{m_t^2} 
-2c_4\lambda_{q_L}^2y_t^2-c_5\lambda_{t_R}^2y_t^2
 \right)
~,
\label{V15exp}
\eea
where $c_i$ are coefficients of order one.
As in the previous model, in order to obtain $v \ll f$
we need  a  partial cancellation between the different terms of  $\mu_h^2$, 
that can be achieved if   $\lambda_{q_L} \sim \lambda_{t_R}\sim \sqrt{y_t g_{\rho}}$ (and if $c_1$ and $c_2$ have opposite sign).  
In this case
the dominant terms in the quartic Higgs couplings are expected to be those proportional to $c_{1,2}^{(2)}\sim1$, that lead to a Higgs mass 
\begin{equation}
m_h^2\sim \frac{N_c}{4\pi^2} \  y_t^2 \  \xi \ m_\rho^2 
= (170\ {\rm GeV})^2 \ \frac{\xi}{0.1}  \left(\frac{m_\rho}{2\ \rm TeV}\right)^2
~.
\end{equation}
In summary, EWSB can be  smoothly obtained when $t_R$ is embedded in the  ${\bf 15}$, with a Higgs naturally close to the presently favoured low-mass region.

%%%%%%%%%%%%%%%%%%%%%%%%%%%%%%%%%%%%%
\section{Dark matter annihilation cross sections} \label{app:A}
%%%%%%%%%%%%%%%%%%%%%%%%%%%%%%%%%%%%%

In this appendix we outline the calculation of the relic density 
and collect for completeness 
all relevant cross sections. 
Following the standard recipe \cite{Gondolo:1990dk}, the central ingredient in the calculation of the DM relic density 
is the thermal-average of its total annihilation cross section times the relative velocity  $v_{rel}$,
\begin{equation}\label{eq:ThermalCrossSection}
\langle \sigma v_{rel}\rangle=\frac{m_{\eta}}{64\pi^4x\,n_{\rm EQ}^2}\int_{4m_{\eta}^2}^{\infty}ds\,\hat{\sigma}(s)\sqrt{s-4m_{\eta}^2}\,\mathcal{K}_{1}\left(x\sqrt{s}/m_{\eta}\right)~,
\end{equation}
where $\hat{\sigma}(s)\equiv s v_{rel}\sum_F\sigma(\eta\eta\to F)$, being $\sigma(\eta\eta\to F)$ the annihilation cross section into the SM final state $F$. In Eq. (\ref{eq:ThermalCrossSection}) the equilibrium number density for $\eta$ is given by
\begin{equation}\label{eq:EquilibriumDensity}
n_{\rm EQ}=\frac{m_{\eta}^3\mathcal{K}_2(x)}{2\pi^2x}~.
\end{equation}
Here $\mathcal{K}_{1,2}$ are the modified Bessel function
and $x\equiv m_{\eta}/T$, where $T$ is the temperature.
A Boltzmann equation describes the evolution of the number density $n(x)$ of the DM particle during the expansion of the Universe,
\begin{equation}\label{eq:BoltzmannEquation}
\frac{d{\rm Y}}{dx}=-\frac{xs(x)}{H}\langle  \sigma v_{rel} \rangle\left({\rm Y}^2-{\rm Y}_{\rm EQ}^2\right)~,
\end{equation}
where the yield is defined by ${\rm Y}\equiv n(x)/s(x)$, and the entropy density $s(x)$ is given by
\begin{equation}\label{eq:Entropy}
s(x)=\frac{2\pi^2g_{*}m_{\eta}^3}{45 x^3}~.
\end{equation}
The Hubble parameter evaluated at $x=1$ is given by
\begin{equation}\label{eq:Hubble}
H=\sqrt{\frac{4\pi^3g_{*}}{45}}\frac{m_{\eta}^2}{M_{\rm PL}}~,
\end{equation}
the value of the Planck mass is $M_{\rm PL}=1.22\cdot 10^{19}$ GeV, and $g_{*}$ is the effective number of degrees of freedom.
The integration of  Eq.~(\ref{eq:BoltzmannEquation}), from $x=0$ to the present  value $x_0=m_{\eta}/T_0$, gives the DM yield
today, ${\rm Y}_0$, 
which in turns is related to the DM relic density \cite{Komatsu:2010fb},
\begin{equation}\label{eq:RelicDensity}
\Omega_{\eta}h^2=\frac{2.74\cdot 10^8m_{\eta}{\rm Y}_{0}}{{\rm GeV}}= 0.1126\pm 0.0036~.
\end{equation}

The computation of the annihilation cross sections makes use of the Feynman rules collected in Table \ref{fig:FRulesCDM}.
The relevant final states are the heavy fermions (the $t$, $b$, $c$ quarks, as well as the $\tau$ lepton), 
the electroweak gauge bosons and the Higgs.
Defining $\hat{ \sigma}_{\eta\eta}(F) \equiv s v_{rel}\sigma(\eta\eta\to F)$, we find
\begin{eqnarray}
\hat{\sigma}_{\eta\eta}(\psi\overline{\psi})|_{\psi_R \in {\bf 6}} &=&
\frac{N_c \, m_\psi^2 \sqrt{1-4m_\psi^2/s}}{\pi f^4}{\biggl\{}
s {\rm Im}(c_{\psi})^2 + (s-4m_{\psi}^2)
\nonumber\\
&\times&  \left.
\left[{\rm Re}(c_{\psi})^2 +
\frac{\Gamma(1-2\xi)^2}{4(1-\xi)^2} +
{\rm Re}(c_{\psi}) \frac{(s-m_h^2)\left[s-2\lambda
f^2(1-\xi)\right](1-2\xi)}{\left[(s-m_h^2)^2+\Gamma_h^2m_h^2\right](1-\xi)}\right]
\right\}~,\nonumber\\ \label{eq:AnnihilationBottom}\\
\hat{\sigma}_{\eta\eta}(\psi\overline{\psi})|_{\psi_R \in {\bf
15}}&=&\frac{N_c m_{\psi}^2
(s-4m_\psi^2)^{3/2}\Gamma}{4\pi  f^4\sqrt{s}}~,\\
\hat{\sigma}_{\eta\eta}(WW)&=&\frac{s^2\,\Gamma}{8\pi f^4}
\left(1-\frac{4m_W^2}{s}+\frac{12m_W^4}{s^2}\right)\sqrt{1-\frac{4m_W^2}{s}}~,\\
\hat{\sigma}_{\eta\eta}(ZZ)&=&\frac{s^2\,\Gamma}{16\pi f^4}
\left(1-\frac{4m_Z^2}{s}+\frac{12m_Z^4}{s^2}\right)\sqrt{1-\frac{4m_Z^2}{s}}~,\label{eq:AnnihilationZ}\\
\hat{\sigma}_{\eta\eta}(hh)&=&\frac{\sqrt{1-4m_{h}^2/s}}{
16f^8\pi(1-\xi^2)^2(s-m_h^2)\left[m_h^4+m_{\eta}^2(s-4m_{h}^2)\right]
}\nonumber\\
&\times&\left\{
\frac{8f^2\xi\left[m_h^4+m_{\eta}^2(s-4m_{h}^2)\right]
\,{\rm arctanh}\left(\Delta\right)\left[m_h^2-2\lambda f^2(1-\xi)\right]^2}
{(s-2m_h^2)^2\Delta}\,A_{\eta\eta}+B_{\eta\eta}
\right\}~,\label{eq:AnnihilationH}\nonumber\\
\end{eqnarray}
where we defined the ratios 
\begin{equation}\label{eq:Ratio}
\Delta\equiv \frac{\sqrt{s-4m_{\eta}^2}\sqrt{s-4m_h^2}}{s-2m_h^2}~,\hspace{5mm} 
\Gamma\equiv
\frac{\left[s-2\lambda f^2(1-\xi)\right]^2}{(s-m_h^2)^2+\Gamma_h^2m_h^2}~.
\end{equation}
The lengthy expressions for the functions $A_{\eta\eta}$ and $B_{\eta\eta}$ 
are listed at the end of this appendix.

Notice  that taking the limit $f\to \infty$, we recover the known expressions for the annihilation  cross sections in the 
non-composite case,
\begin{eqnarray}
\left.\hat{\sigma}_{\rm \eta\eta}(\psi\overline{\psi})\right|_{f\to \infty}&=&\frac{N_c\, \lambda^2m_\psi^2
(s-4m_\psi^2)^{3/2}}{\pi\sqrt{s}[(s-m_h^2)^2+\Gamma_h^2m_h^2]}~,\label{eq:AnnihilationSSBottom}\\
\left.\hat{\sigma}_{\rm \eta\eta}(WW)\right|_{f\to \infty}&=&\frac{s^2\lambda^2}{2\pi\left[(s-m_h^2)^2+\Gamma_h^2m_h^2\right]}
\left(1-\frac{4m_W^2}{s}+\frac{12m_W^4}{s^2}\right)\sqrt{1-\frac{4m_W^2}{s}}~,\\
\left.\hat{\sigma}_{\rm \eta\eta}(ZZ)\right|_{f\to \infty}&=&\frac{s^2\lambda^2}{4\pi\left[(s-m_h^2)^2+\Gamma_h^2m_h^2\right]}
\left(1-\frac{4m_Z^2}{s}+\frac{12m_Z^4}{s^2}\right)\sqrt{1-\frac{4m_Z^2}{s}}~,\label{eq:AnnihilationSSZ}\\
\left.\hat{\sigma}_{\rm \eta\eta}(hh)\right|_{f\to \infty}&=&\frac{\lambda^2\sqrt{1-4m_h^2/s}}{4\pi}
\left(\mathcal{A}_{\rm \eta\eta}+\mathcal{B}_{\rm \eta\eta}\,v^2+\mathcal{C}_{\rm \eta\eta}\,v^4\right)~,\label{eq:AnnihilationSSH}
\end{eqnarray}
where
\begin{eqnarray}
\mathcal{A}_{\rm \eta\eta}&=&\frac{(s+2m_h^2)^2}{(s-m_h^2)^2}~,\\
\mathcal{B}_{\rm \eta\eta}&=&-\frac{
16\lambda(s+2m_h^2)\,{\rm arctanh}\left(\Delta\right)
}{
\Delta(s-2m_h^2)(s-m_h^2)
}~,\\
\mathcal{C}_{\rm \eta\eta}&=&32\lambda^2\left[
\frac{
{\rm arctanh\left(\Delta\right)}
}{
\Delta(s-2m_h^2)^2
}+\frac{1}{4m_{\eta}^2(s-4m_h^2)+4m_{h}^4}
\right]~.
\end{eqnarray}

We plot the annihilation cross sections in Fig.~\ref{fig:SingoliCanali}, as a function of the DM mass, for 
$v_{rel}=2/3$, $\lambda=10^{-2}$ and $m_h=125$ GeV, taking $f=500$ GeV in the left-hand panel  and $f=\infty$
in the right-hand panel. 
For comparison we also show (black dashed line) the typical value for $\langle \sigma v_{rel}\rangle$ suggested by 
Eq.~(\ref{eq:GoldenRule}), needed for the thermal freeze-out of the DM particle.
In the left panel of Fig.~\ref{fig:SingoliCanali} one can distinguish the four DM mass regions
described  in section \ref{Sec:RelicDensity}, 
that characterize the qualitative behavior of the composite singlet relic density.

\begin{figure}[!tb!]
  \begin{minipage}{0.4\textwidth}
   \centering
   \includegraphics[scale=0.65]{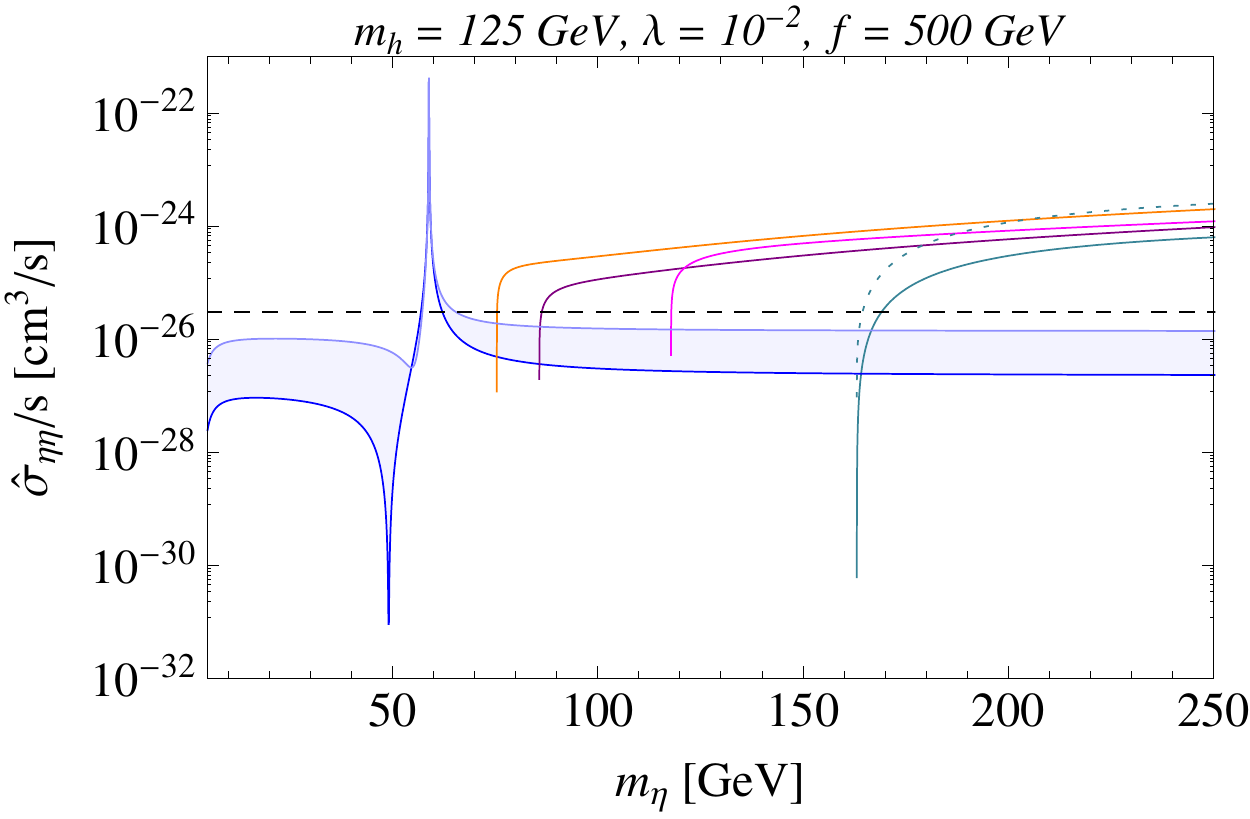}
     \end{minipage}\hspace{1.5 cm}
   \begin{minipage}{0.4\textwidth}
    \centering
    \includegraphics[scale=0.65]{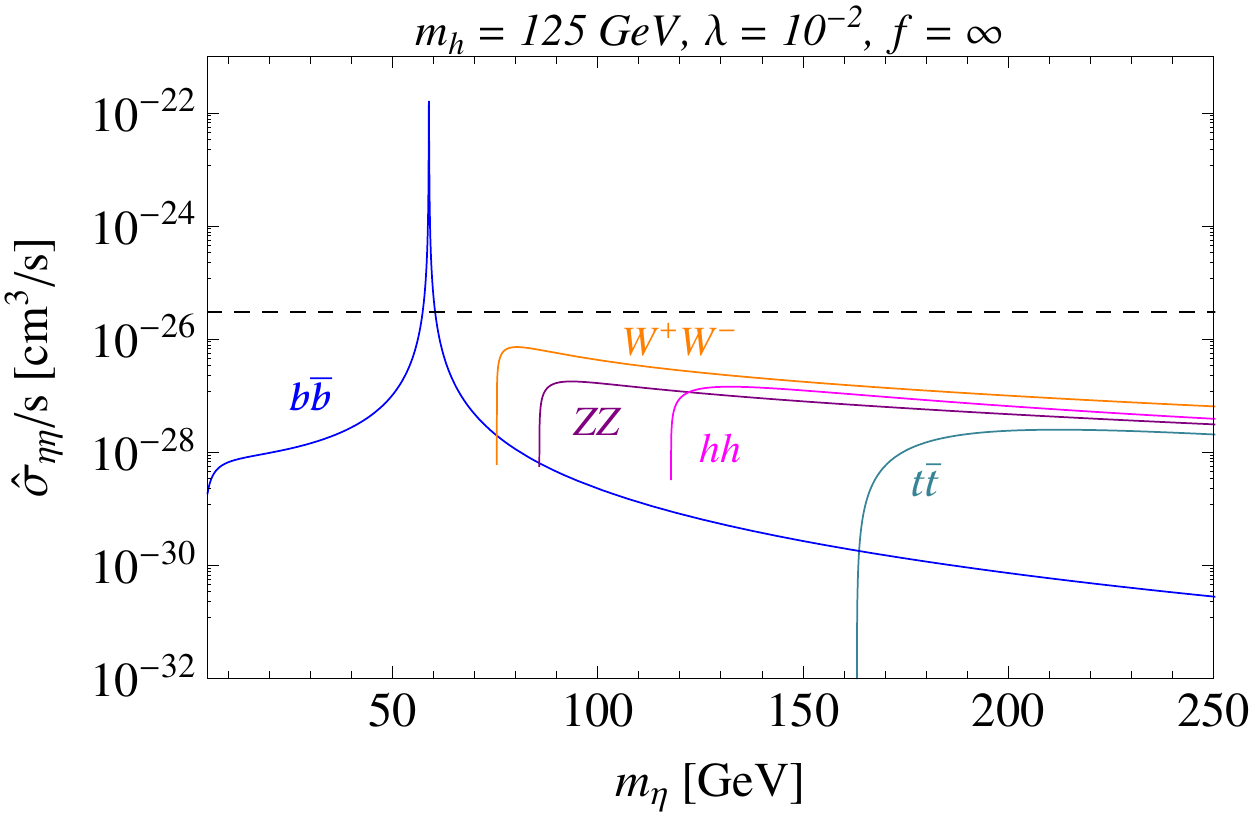}
        \end{minipage}
 \caption{\emph{Annihilation cross sections times the relative velocity $v_{rel}$ for the DM candidate $\eta$. We plot 
 Eqs.~(\ref{eq:AnnihilationBottom})-(\ref{eq:AnnihilationH}) for $f=500$ GeV (left-hand panel), and 
 Eqs.~(\ref{eq:AnnihilationSSBottom})-(\ref{eq:AnnihilationSSH}) for the non-composite case (right-hand panel); in both cases $\lambda=10^{-2}$, $m_h=125$ GeV, $v_{rel}=2/3$.
We vary the coefficient $c_b$ as in Fig.~\ref{fig:BoltzmannRelic}.
 For the top quark channel, we show both Case 1 and 2 (dotted and solid cyan lines, respectively). 
 The black dashed line indicates the benchmark value for DM freeze-out, $\langle \sigma v_{rel}\rangle = 3\cdot 10^{-26}$ cm$^{3}$s$^{-1}$.}}
 \label{fig:SingoliCanali}
\end{figure}
%%%%%%%%%%%%%%%%%%%%%%%%%%%%%%%%%%%%%%%%%%%%

Finally we report the expressions for the coefficients $A_{\eta\eta}$ and $B_{\eta\eta}$ in Eq. (\ref{eq:AnnihilationH}):

\begin{eqnarray}
A_{\eta\eta}&=& f^2 \left\{m_h^4 \left[4 f^2 \lambda (\xi -1) (3 \xi -2)+(5 \xi -4) s\right]-m_h^2 \xi  \left[s+2 f^2 \lambda(1-\xi)\right]^2\right.\nonumber\\&+&\left.
(1-\xi) s \left[4 f^4 \lambda^2 (1-\xi) \xi -2 f^2 \lambda (1-2 \xi) s+s^2\right]+3 m_h^6 \xi\frac{}{}\right\}~,\\
B_{\eta\eta}&=&\frac{f^4}{s-m_{h}^2}\,\,\sum_{n=0}^{6} g_{2n}\,m_h^{2n}~,
\end{eqnarray}
with
\begin{eqnarray}
g_0 &=&
 (1-\xi)^2 s^2 \left\{32 f^8 \lambda^4 (1-\xi)^2 \xi ^2+m_{\eta}^2 s \left[2 f^2 \lambda(2 \xi -1)+s\right]^2\right\}~,\\
 g_2&=& 2 s (\xi -1) \xi  \left\{m_{\eta}^2 s^2 \left[2 f^2\lambda (2 \xi -1)+s\right]+32 f^6 \lambda^3 (\xi -1)^2 \xi  \left[s-f^2 \lambda (\xi -1)\right]\right\}~,\nonumber\\
 \\
 g_4&=&  (\xi -1)^2 \left\{32 f^8 \lambda^4 (\xi -1)^2 \xi ^2-128 f^6 \lambda^3 (\xi -1) \xi ^2 s+4 f^4 \lambda^2 [4 \xi  (4 \xi -1)+1] s^2+\right.\nonumber\\
&& \left.4 f^2 \lambda (2 \xi -1) s^3+s^4\right\}-m_{\eta}^2 s \left\{48 f^4 \lambda^2 \left(2 \xi ^2-3 \xi +1\right)^2+\right.\nonumber\\&&\left.48 f^2\lambda(\xi -1)^2 (2 \xi -1) s+[\xi  (11 \xi -24)+12] s^2\frac{}{}\right\}~,\\
g_6&=& 2 (\xi -1) \left\{32 f^6 \lambda^3 (\xi -1)^2 \xi ^2-8 f^4 \lambda^2 (\xi -1) [2 \xi  (\xi +2)-1] s-\right.\nonumber\\
&&\left.4 m_{\eta}^2 \left[2 f^2 \lambda(2 \xi -1)+s\right] \left[4 f^2 \lambda (\xi -1) (2 \xi -1)+(5 \xi -2) s\right]+\right.\nonumber\\&&\left.2 f^2\lambda[\xi  (14 \xi -13)+4] s^2+(3 \xi -2) s^3\right\}~,\\
g_8&=&16 f^2 \lambda (\xi -1) \left\{f^2 \lambda (\xi -1) [\xi  (7 \xi -4)+1]+4 m_{\eta}^2 (1-2 \xi ) \xi \right\}+\nonumber\\
&& 4 s \left\{4 f^2\lambda (\xi -1) [2 (\xi -2) \xi +1]+m_{\eta}^2 (8-11 \xi ) \xi \right\}+(3 \xi -2) (5 \xi -2) s^2~,\\
g_{10}&=& 8 \xi  \left[2 f^2 \lambda (\xi -1) (3 \xi -1)-2m_{\eta}^2 \xi +(\xi -1) s\right]~,\\
g_{12}&=& 6\xi^2~.
\end{eqnarray}

%%%%%%%%%%%%%%%%%%%%%%%%%%%%%%%%%%%%%
\section{Dark matter scattering cross section on nuclei}
\label{app:B}
%%%%%%%%%%%%%%%%%%%%%%%%%%%%%%%%%%%%%

The spin-independent elastic scattering cross section of a real scalar $\eta$ on a nucleus can be parameterized as
\begin{equation}\label{eq:DD1}
\sigma_{\rm SI}=\frac{1}{\pi} 
\left(\frac{m_{\eta}m_n}{m_{\eta}+m_n}\right)^2
\frac{\left[Zf_p+(A-Z)f_n \right]^2}{A^2}~,
\end{equation}
where $m_n$ is the neutron mass, $Z$ and $A-Z$ are the number of protons and neutrons in the nucleus, and
the function $f_p$ ($f_n$) describes the coupling between $\eta$ and protons (neutrons): 
\begin{eqnarray}\label{DMcouplingsAPP}
f_{n,p} &=&  \sum_{q=u,d,s}f^{(n,p)}_{T_q}\,a_q\,\frac{m_{n,p}}{m_q}+\frac{2}{27}\,f^{(n,p)}_{T_{G}}\sum_{q=c,b,t}\,a_q\,\frac{m_{n,p}}{m_q}~.\label{eq:DD2}
\end{eqnarray}
The hadron matrix elements $f^{(n,p)}_{T_q}$ parametrize the quark content of the nucleon.
They are extracted from pion-nucleon diffusion measurements. Following a recent analysis \cite{Cheng:2012qr} we take
\begin{equation}
\begin{array}{ccc}\vspace{2mm}
  f_{T_u}^{(p)}=0.017~,    & f_{T_d}^{(p)}=0.022~, & f_{T_s}^{(p)}=0.053~, \\ 
   f_{T_u}^{(n)}=0.011~,    & f_{T_d}^{(n)}=0.034~,  & f_{T_s}^{(n)}=0.053~. 
\end{array}
\end{equation}
The DM couples also to the gluons in the nucleus, through a loop of heavy quarks (those with $m_q \gg \Lambda_{QCD}$).
The corresponding matrix element is given by
\begin{equation}\label{eq:GluonsCoupling}
f_{T_{G}}^{(n,p)}=1-\sum_{q=u,d,s}f_{T_q}^{(n,p)}~.
\end{equation}
In Eq.  (\ref{eq:DD2}), the parameter $a_q$ describes the short-distance effective interactions between $\eta$ and a given quark,
normalized as
\begin{equation}\label{eq:ScalarEffectiveInteractions}
\mathcal{L}_{\eta}\supset\sum_q a_q\eta^2\overline{q}q~.
\end{equation}

In our composite model there are two sources for this effective interaction: the direct coupling $c_q$ of $\eta$ to quarks, and a coupling mediated by the exchange of the Higgs boson.
More precisely, only Re$(c_q)$ enters in $a_q$, and thus contributes to  $\sigma_{\rm SI}$, while Im$(c_q)$
leads to a coupling $\eta^2 \bar q \gamma_5 q$, that contributes only to the spin-dependent cross section.
Coming to the Higgs in the t-channel,
one should notice that the momentum transfer $q^2$ in the elastic scattering is very small,
therefore the Higgs-exchange reduces to a contact interaction suppressed by $m_h^2$. 
In addition, the derivative coupling between $\eta$ and $h$ can be neglected, because it is
proportional to $q^2$, and one is left with the coupling $\lambda$ only.
All in all, we find
\begin{eqnarray}
a_q|_{q_R \in {\bf 6}}&=& \frac{m_q}{m_{\eta}}\left[\frac{{\rm
Re}(c_q)}{f^2}+\frac{\lambda(1-2\xi)}{m_h^2}\right]~,\label{eq:Direct1}\\
a_{q}|_{q_R \in {\bf 15}}  &=&\frac{m_q}{m_\eta}\
\frac{\lambda(1-\xi)}{m_h^2}~.
\end{eqnarray}
Putting all the ingredients together, one recovers the numerical approximation shown in \eq{SIApprox2}.

%%%%%%%%%%%%%%%%%%
%%%%%%%%%%%%%%%%%%

\bibliographystyle{JHEP}

\end{document}